\documentclass{aa}

\usepackage[T1]{fontenc}
\usepackage[latin1]{inputenc}
\usepackage{wasysym}
\usepackage{graphicx}
\usepackage{natbib}
\usepackage{txfonts}

\newcommand{\eq}[1]{\begin{equation}  #1 \end{equation}}

\newcommand{\eqa}[1]{\begin{eqnarray}   #1 \end{eqnarray}}

\newcommand{\br}[1]{\left( #1 \right)}
\newcommand{\bc}[1]{\left\{ #1 \right\}}
\newcommand{\bb}[1]{\left[ #1 \right]}
\newcommand{\ba}[1]{\left\langle #1 \right\rangle}

\newcommand{\nn}{\nonumber}

\newcommand{\dd}{{\rm d}}
\newcommand{\expo}[1]{~{\rm e}^{ #1 }}
\newcommand{\vek}[1]{\mbox{\boldmath $#1$}}
\newcommand{\svek}[1]{\mbox{\boldmath \scriptsize $#1$}}  
\newcommand{\ic}{{\rm i}}

\begin{document}

\title{The removal of shear-ellipticity correlations\\ from the cosmic shear signal}
\subtitle{Influence of photometric redshift errors on the nulling technique}

\author{B. Joachimi \and P. Schneider}

\offprints{B. Joachimi,\\
    \email{joachimi@astro.uni-bonn.de}
}

\institute{Argelander-Institut f\"ur Astronomie (AIfA), Universit\"at Bonn, Auf dem H\"ugel 71, 53121 Bonn, Germany}

\date{Received 4 May 2009 / Accepted 20 August 2009}

\abstract{}
{Cosmic shear, the gravitational lensing on cosmological scales, is regarded as one of the most powerful probes for revealing the properties of dark matter and dark energy. To fully utilize its potential, one has to be able to control systematic effects down to below the level of the statistical parameter errors. Particularly worrisome in this respect is the intrinsic alignment of galaxies, causing considerable parameter biases via correlations between the intrinsic ellipticities of galaxies and the gravitational shear, which mimic lensing. Since our understanding of the underlying processes of intrinsic alignment is still poor, purely geometrical methods are required to control this systematic. In an earlier work we proposed a nulling technique that downweights this systematic, only making use of its well-known redshift dependence. We assess the practicability of nulling, given realistic conditions on photometric redshift information.}
{For several simplified intrinsic alignment models and a wide range of photometric redshift characteristics, we calculate an average bias before and after nulling. Modifications of the technique are introduced to optimize the bias removal and minimize the information loss by nulling. We demonstrate that one of the presented versions of nulling is close to optimal in terms of bias removal, given the high quality of photometric redshifts. 
Although the nulling weights depend on cosmology, being composed of comoving distances, we show that the technique is robust against an incorrect choice of cosmological parameters when calculating the weights. 
Moreover, general aspects such as the behavior of the Fisher matrix under parameter-dependent transformations and the range of validity of the bias formalism are discussed in an appendix.}
{Given excellent photometric redshift information, i.e. at least 10 bins with a dispersion $\sigma_{\rm ph} \lesssim 0.03$, a negligible fraction of catastrophic outliers, and precise knowledge about the bin-wise redshift distributions as characterized by a scatter of 0.001 or less on the median redshifts, one version of nulling is capable of reducing the shear-intrinsic ellipticity contamination by at least a factor of 100. Alternatively, we describe a robust nulling variant which suppresses the systematic signal by about 10 for a very broad range of photometric redshift configurations, provided basic information about $\sigma_{\rm ph}$ in each of $\gtrsim$ 10 photometric redshift bins is available. Irrespective of the photometric redshift quality, a loss of statistical power is inherent to nulling, which amounts to a decrease of the order $50\,\%$ in terms of our figure of merit under conservative assumptions.
}
{}


\maketitle

\section{Introduction}
\label{sec:introduction}

Within a few years only cosmic shear, the weak gravitational lensing of distant galaxies by the large-scale structure of the Universe, has evolved from its first detections \citep{bacon00,kaiser00,vwaer00,wittman00} into one of the most promising methods for shedding light on cosmological issues in the near future \citep{albrecht06,peacock06}. Probing both the geometry of the Universe and the formation of structure, cosmic shear is able to put tight constraints on the parameters of the cosmological standard model and its extensions, breaking degeneracies when combined with other methods such as the cosmic microwave background, baryonic acoustic oscillations, galaxy redshift surveys, and supernova distance measurements \citep[e.g.][]{hu02b,spergel07}. This way, questions of fundamental physics concerning the nature of dark matter and dark energy \citep[see e.g.][]{schaefer09} and the law of gravity \citep[e.g.][]{thomas08} can be answered.

While recent observations have already been able to decrease statistical errors considerably \citep[see e.g.][]{jarvis06,hoekstra06,semboloni06,hetterscheidt07,benjamin07,fu07}, planned surveys with instruments like Euclid, JDEM, LSST, or SKA will provide weak lensing data with unprecedented precision. The anticipated high quality of data enforces a careful and complete treatment of systematic errors, which has become one focus of current work in the field --  consider for instance \citet{step1}, \citet{step2}, and \citet{great08} regarding galaxy shape measurements.

A potentially serious systematic to cosmic shear measurements is the intrinsic alignment of galaxies, a physical alignment of galaxies that can mimic the apparent shape alignment of galaxy images induced by gravitational lensing. At the two-point level, all measures of cosmic shear are based on correlators between the measured ellipticities $\epsilon$ of galaxies, where $\epsilon$ is a complex number, coding the absolute value of the ellipticity and the orientation of the galaxy image with respect to a reference axis. In the approximation of weak lensing $\epsilon$ can be written as the sum of the intrinsic ellipticity $\epsilon^{\rm s}$ of the galaxy and the gravitational shear $\gamma$. Applying this relation, the correlator of ellipticities for two galaxy populations $i$ and $j$ reads
\eqa{
\label{eq:epscorrelators}
\ba{\epsilon_i \epsilon_j^*} &=& \underbrace{\ba{\gamma_i \gamma_j^*}}+\underbrace{\ba{\epsilon_i^{\rm s} \epsilon_j^{{\rm s}*}}}+\underbrace{\ba{\gamma_i \epsilon_j^{{\rm s}*}}+\ba{\epsilon_i^{\rm s} \gamma_j^*}}\;.\\ \nn
&& \hspace*{0.2cm} {\rm GG} \hspace*{0.9cm} {\rm II} \hspace*{1.6cm} {\rm GI}
}
If one assumes that the intrinsic ellipticities of galaxies are randomly oriented in the sky, only the desired lensing (GG) term remains on the right-hand side. However, when galaxies are subject to the tidal forces of the same matter structure, their shapes can intrinsically align and become correlated, thus causing a non-vanishing II term. Moreover, a matter overdensity can align a close-by galaxy and at the same time contribute to the lensing signal of a background object, which results in non-zero correlations between gravitational shear and intrinsic ellipticities or a GI term \citep[HS04 hereafter]{hirata04}.

The alignment of dark matter haloes, resulting from external tidal forces, has been subject to extensive study, both analytic and numerical (\citealp{croft00,heavens00,lee00,catelan01,crittenden01,jing02,mackey02}; HS04; \citealp{bridle07a,schneiderm09}). The galaxies in turn are assumed to align with the angular momentum vector (in the case of spiral galaxies) or the shape of their host halo (in the case of elliptical galaxies), which is suggested by the observed correlations of galaxy spins \citep[e.g.][]{pen00} and galaxy ellipticities \citep[e.g.][]{brainerd09}. However, this alignment is not perfect -- see for instance \citet{bosch02}, \citet{okumura08}, and \citet{okumura09}. The intrinsic correlations of galaxy properties cause non-zero II and GI signals, as observationally verified in several surveys by e.g. \citet{brown02}, \citet{heymans04}, \citet{mandelbaum06}, \citet{hirata07}, and \citet{brainerd09}. 

Observations as well as predictions from theory are consistent with a contamination of the order of 10$\,\%$ by both II and GI signal for future cosmic shear surveys, which makes the control of these systematics crucial. However, analytic progress to calculate intrinsic alignment correlations beyond linear theory is cumbersome, and the inclusion of gas physics to fully simulate the formation and evolution of galaxies in their dark matter haloes is computationally still too expensive (see e.g. \citealp{schaefer08} for a review on the work about galaxy spin correlations), so that for the time being our understanding of intrinsic alignment remains at the level of toy models. 

Hence, removal techniques should rely on intrinsic alignment models as little as possible. The II signal is relatively straightforward to eliminate because it is restricted to pairs of galaxies that are physically close to each other, both galaxies being affected by the same matter structure \citep{king02,king03,heymans03,takada04b}. For an application of the II removal to the COMBO-17 survey see \citet{heymans04}. 

First ideas how to control the GI signal were already put forward by HS04. \citet{king05} uses a set of template functions to fit the lensing and intrinsic alignment signals simultaneously, making use of their different dependence on angular scales and redshift. Similarly, \citet{bridle07} investigate the effect of the GI term on parameter constraints by binning the systematic signal in angular frequency and redshift with free parameters, which are then marginalized over. In both approaches an intrinsic alignment toy model is used as fiducial model. Increasing freedom in the representation of the GI signal is achieved at the cost of a bigger number of nuisance parameters, which dilutes the cosmological information that can be extracted from the data.

In addition to ellipticity correlations one can also measure galaxy densities in cosmic shear surveys, so that ellipticity-density and density-density correlations can be added to the data analysis. This information is then used to self-calibrate systematic effects of weak lensing \citep[e.g.][]{hu03,bernstein08}. \citet{zhang08} applies the self-calibration technique to the GI contamination, deriving an approximate relation between GI and the galaxy density-intrinsic ellipticity correlations.

In a purely geometric approach \citet{joachimi08b}, JS08 hereafter, have presented a technique to null the GI signal, based exclusively on weak lensing data. Making use of the characteristic dependence on redshift, new cosmic shear measures are constructed that are completely free of any possible GI systematic, given perfect redshift information. In a case study it was shown in JS08 that for more than about 10 redshift bins up to $z=4$, still without photometric redshift errors, the nulling technique only moderately widens parameter constraints. To demonstrate its practicability, it is vital to assess the performance of nulling in presence of photometric redshift inaccuracies and to quantify the actual suppression of the GI signal since the removal is not necessarily perfect as idealized assumptions in the derivation of the method have been made. It is the scope of this work to investigate the modification of statistical and systematic errors by the nulling technique in a more realistic setup, including photometric redshift errors. Furthermore, we are going to provide minimum requirements on the quality of redshift information to be able to practically apply nulling.

The paper is structured as follows: In Sect.$\,$\ref{sec:method} we review the nulling technique, slightly modifying the approach to further simplify notation and usage. Moreover, we give an overview on the Fisher matrix and bias formalism in the context of the data transformation that corresponds to nulling. Section$\,$\ref{sec:modelling} summarizes our model specifications concerning photometric redshift errors, lensing data, and intrinsic alignment signals. We determine the nulling parameters such that the corresponding transformation removes a maximum of systematic signal in Sect.$\,$\ref{sec:optimization}. Besides, we address the dependence of the nulling weights on cosmology. In Sect.$\,$\ref{sec:zinfo} the performance of nulling in terms of photometric redshift binning is elaborated on, leading to considerations of the minimum information loss of this technique. In addition, we develop a weighting scheme to control intrinsic alignment contamination, not eliminated by nulling itself. Section \ref{sec:photozerrors} deals with the effect of photometric redshift uncertainty and assesses to what extent the chosen nulling versions are optimal. The influence of catastrophic outliers in and of uncertainty in the parameters of the redshift distributions is quantified in Sect.$\,$\ref{sec:furthercharacteristics}. In Sect.$\,$\ref{sec:conclusions} we summarize our findings and conclude. The appendices provide a discussion of parameter-dependent transformations of the Fisher matrix and a formal derivation of the bias formalism, including an assessment of its validity.

\section{Method}
\label{sec:method}

\subsection{Nulling technique}
\label{sec:nulling}

We briefly review the principles of the nulling technique as presented in JS08 and develop a compact formalism. As before, we restrict our considerations to Fourier space by using power spectra as the cosmic shear measures, but it is straightforward to implement the formalism in terms of any of the second-order real-space measures. Throughout the paper a spatially flat universe is assumed. For recent reviews on weak lensing see e.g. \citet{munshi08} for theoretical issues and \citet{hoekstra08} who focus on observational aspects; \citet{heavens08} provides a concise overview. We largely follow the notation of \citet{Schneider06}.

Consider a cosmic shear survey that is divided into $N_z$ redshift slices by means of photometric redshift information, yielding a data set of tomography convergence power spectra $P_{\rm GG}^{(ij)}(\ell)$, where the indices $i$ and $j$ run from 1 to $N_z$, and where the angular frequency $\ell$ denotes the Fourier variable on the sky. We use the convention that in the superscript of the power spectra the first bin refers to the redshift distribution with lower median redshift, i.e. $i \leq j$. The convergence power spectra are radial projections of the three-dimensional power spectrum of matter density fluctuations $P_{\delta\delta}$ as given by Limber's equation in Fourier space \citep{kaiser92},
\eq{
\label{eq:limber}
P_{\rm GG}^{(ij)}(\ell) = \frac{9H_0^4 \Omega_{\rm m}^2}{4 c^4} \!\!\! \int^{\chi_{\rm hor}}_0 \!\!\!\! \dd \chi\; g^{(i)}(\chi)\, g^{(j)}(\chi)  \bc{1+z(\chi)}^2 P_{\delta\delta} \br{\frac{\ell}{\chi},\chi}\;.
}
Here and in the following, the dependence of the power spectra on time is encoded in the second argument, respectively. The redshift is denoted by $z$, while $\chi$ is the comoving distance, with its maximum at the comoving horizon distance $\chi_{\rm hor}$. These two quantities are related via the distance-redshift relation
\eq{
\label{eq:wz}
\chi(z)=\frac{c}{H_0}\int^z_0 {\rm d}z' \bc{ \Omega_{\rm m} (1+z')^3 + \Omega_{\rm DE}(z') }^{-1/2}\;,
}
where $\Omega_{\rm DE}(z) \equiv \Omega_{{\rm DE},0}$ in case of a cosmological constant. The parametrization of $\Omega_{\rm DE}(z)$ in a universe with variable dark energy is given in Sect.$\,$\ref{sec:ps}. The weighting in the projection (\ref{eq:limber}), specific to weak gravitational lensing, is the lensing efficiency 
\eq{
\label{eq:lenseff}
g^{(i)}(\chi) = \int_\chi^{\chi_{\rm hor}} \dd \chi'\, p^{(i)}(\chi')\, \br{1 - \frac{\chi}{\chi'}}\;,
}
where $p^{(i)}(\chi)$ is the normalized probability distribution of comoving distances of a galaxy population $i$. Hence, the lensing efficiency corresponds to the ratio $D_{\rm ds}/D_{\rm s}$ of the angular diameter distance between lens and source and the one between observer and source, averaged over the source distances of the galaxy population $i$.

Intrinsic alignment leads to correlations between the intrinsic ellipticities of galaxies and between intrinsic ellipticity and gravitational shear, thereby adding a systematic signal to the lensing observables (\ref{eq:limber}). In analogy to (\ref{eq:limber}), the II and GI power spectra can be written as (HS04)
\eqa{
\label{eq:limberII}
P_{\rm II}^{(ij)}(\ell) &=& \int^{\chi_{\rm hor}}_0 \dd \chi\; p^{(i)}(\chi)~p^{(j)}(\chi)\; \chi^{-2} P_{\gamma^{\rm I} \gamma^{\rm I}} \br{\frac{\ell}{\chi},\chi}\;;\\ \nn
P_{\rm GI}^{(ij)}(\ell) &=& \frac{3H_0^2 \Omega_{\rm m}}{2 c^2} \int^{\chi_{\rm hor}}_0 \dd \chi\; \br{p^{(i)}(\chi)~g^{(j)}(\chi) + g^{(i)}(\chi)~p^{(j)}(\chi)}\\
\label{eq:limberGI}
&& \times\; \bc{1+z(\chi)}\; \chi^{-1} P_{\delta \gamma^{\rm I}} \br{\frac{\ell}{\chi},\chi}\;.
}
In order to define the three-dimensional power spectra employed here, we write $\epsilon^{\rm s}=\gamma^{\rm I}+\epsilon^{\rm rnd}$, i.e. the intrinsic ellipticity is split up into the contributions by an intrinsic shear field $\gamma^{\rm I}(\vek{x})$ that contains the intrinsic alignment effects, being continuous as a function of position vector $\vek{x}$, and a purely random component $\epsilon^{\rm rnd}$. The latter term is correlated neither with gravitational or intrinsic shear, nor with $\epsilon^{\rm rnd}$ of other galaxies. Analogously to the lensing case one can introduce an intrinsic convergence $\kappa^{\rm I}$ such that $\tilde{\kappa}^{\rm I}(\vek{k})=\tilde{\gamma}^{\rm I}(\vek{k}) \expo{-2\ic \varphi_k}$, where the tilde denotes the Fourier transform, and where $\varphi_k$ is the azimuthal angle of the wave vector $\vek{k}$.

Then one defines the intrinsic shear E-mode power spectrum $P_{\gamma^{\rm I} \gamma^{\rm I}}$ and the matter-intrinsic shear cross-power spectrum $P_{\delta \gamma^{\rm I}}$ as
\eqa{
\label{eq:def3DpsII}
\ba{\tilde{\kappa}_{\rm E}^{\rm I}(\vek{k},\chi)~\tilde{\kappa}_{\rm E}^{{\rm I}\,*}(\vek{k'},\chi)} &=& (2\pi)^3 ~\delta^{(3)}_{\rm D}(\vek{k} - \vek{k'}) P_{\gamma^{\rm I} \gamma^{\rm I}}(k,\chi)\;,\\
\label{eq:def3DpsGI}
\ba{\tilde{\delta}(\vek{k},\chi)~\tilde{\kappa}_{\rm E}^{{\rm I}\,*}(\vek{k'},\chi)} &=& (2\pi)^3 ~\delta^{(3)}_{\rm D}(\vek{k} - \vek{k'}) P_{\delta \gamma^{\rm I}}(k,\chi)\;,
}
where $\delta_{\rm D}$ is the Dirac delta-distribution. In analogy to (\ref{eq:def3DpsII}) a B-mode intrinsic shear power spectrum can be defined as well \citep{schneiderm09}. The cross-power spectra between intrinsic shear E- and B-mode $\ba{\tilde{\kappa}_{\rm E}^{\rm I}(\vek{k},\chi)~\tilde{\kappa}_{\rm B}^{{\rm I}\,*}(\vek{k'},\chi)}$ and between matter and intrinsic shear B-mode $\ba{\tilde{\delta}(\vek{k},\chi)~\tilde{\kappa}_{\rm B}^{{\rm I}\,*}(\vek{k'},\chi)}$ should vanish if one demands parity invariance of the intrinsic shear field \citep[see][]{schneider03}. 

To see the equivalence between the definition in (\ref{eq:def3DpsGI}) and the one in HS04, consider the Fourier transform of the correlator $\ba{\delta(0,\chi)\, \gamma_+^{\rm I}(\vek{x},\chi)}$, which is given by
\eqa{
\label{eq:FTcorrelator}
\ba{\delta(0,\chi)\, \gamma_+^{\rm I}(\vek{x},\chi)} &=& \int \frac{\dd^3 k}{(2\,\pi)^3} \int \frac{\dd^3 k'}{(2\,\pi)^3} \expo{- \ic \svek{k} \cdot \svek{x} }\\ \nn
&& \hspace*{1cm} \times \cos(2\varphi_k)\; \ba{\tilde{\delta}(\vek{k'},\chi)\; \tilde{\kappa}_{\rm E}^{\rm I\,*}(\vek{k}, \chi) }\;,
}
where it was assumed that the $+$-component of the intrinsic shear is measured along $\vek{x_\perp}$, the transverse separation component of the position vector $\vek{x}$. Inserting (\ref{eq:def3DpsGI}) and integrating along the line of sight, one obtains 
\eq{
\label{eq:defcrosspower}
\int \dd x_\parallel \ba{\delta(0,\chi)\, \gamma_+^{\rm I}(\vek{x},\chi)} = - \int \frac{\dd k\, k}{2\,\pi} J_2(k x_\perp)\; P_{\delta \gamma^{\rm I}} (k,\chi)\;, 
}
where the definition of the second-order Bessel function of the first kind, written as $J_2$, was employed in addition. By making use of the orthogonality relations of Bessel functions, one arrives at the defining equation of $P_{\delta \gamma^{\rm I}}$ in HS04, Eq.$\,$12. 

Note that HS04 account for source clustering by using the weighted intrinsic shear $\gamma^{\rm I} (1+\delta_{\rm g})$, where $\delta_{\rm g}$ is the density contrast of galaxies. Since in this work we merely implement the linear alignment GI signal, which does not have any contribution due to source clustering, we drop the tilde that marks the weighted intrinsic shear in the notation of HS04 to avoid confusion with Fourier transforms.

The explicit form of both $P_{\gamma^{\rm I} \gamma^{\rm I}}$ and $P_{\delta \gamma^{\rm I}}$ depend on the intricacies of galaxy formation and evolution within their dark matter environment, and are to date only poorly constrained from both theory and observations \citep[for a recent theoretical approach based on the halo model see][]{schneiderm09}. Thus, it is currently impossible to model these systematics with the necessary accuracy to precisely measure cosmological parameters by cosmic shear without risking a severe bias.

Consequently, one has to rely on geometrical methods to remove the intrinsic alignment systematics. The II signal stems from pairs of galaxies that are physically close, i.e. close both on the sky and in (spectroscopic) redshift. As long as the redshift distributions of galaxies are relatively concentrated, one can thus eliminate the II correlations by removing pairs of galaxies close in photometric redshift estimates \citep{king02,heymans03}, as is also evident from the weighting in the integrand of (\ref{eq:limberII}). \citet{takada04b} have shown that excluding the auto-correlations from the analysis increases statistical errors only moderately by about 10$\,\%$ when using at least five redshift slices. We follow this approach by excluding auto-correlations from our investigations. A more sophisticated downweighting scheme of the II signal in presence of tomography cosmic shear data can be readily incorporated into the nulling technique. Hence, we are going to neglect the contamination by the II signal in what follows. However, as we will also deal with cases of large photometric errors, an II signal is expected to be present in cross-correlations of different redshift distributions. This limits the validity of dropping the II signal, as will be assessed in Sect.$\,$\ref{sec:GIsignal}.

To eliminate the GI contamination, we null all contributions to the lensing signal from matter, located at the redshift of the galaxies in distribution $i$, i.e. the distribution with lower median redshift. The derivation of the nulling technique is based on the assumption of narrow photometric redshift bins, so that we write
\eq{
\label{eq:bindelta}
p^{(i)}(\chi) \approx \delta_{\rm D}(\chi - \chi(\hat{z}_i))\;,
}
where $\chi(\hat{z}_i)$ is the comoving distance corresponding to an appropriately chosen redshift $\hat{z}_i$ within distribution $i$. As a consequence, the lensing efficiency (\ref{eq:lenseff}) simplifies to $g^{(i)}(\chi) \approx 1-\chi/\chi(\hat{z}_i)$ for $\chi \leq \chi(\hat{z}_i)$ and 0 else. Introducing a weight function $B^{(i)}(\chi)$, one can define a modified lensing efficiency via
\eq{
\label{eq:modlenseff}
\hat{g}^{(i)}(\chi) \equiv \int_{\chi}^{\chi_{\rm hor}} \dd \chi' ~B^{(i)}(\chi')\, \br{1 - \frac{\chi}{\chi'}}\;,
}
which constitutes a weighted integral over the approximated lensing efficiency. The lower integration limit was changed from 0 to $\chi$ because the lensing efficiency in the integrand vanishes for $\chi' < \chi$, see above. The weight function is constrained by the equation
\eq{
\label{eq:zwangsbed}
\hat{g}^{(i)}(\chi(\hat{z}_i)) = \int_{\chi(\hat{z}_i)}^{\chi_{\rm hor}} \dd \chi' ~B^{(i)}(\chi')\, \br{1 - \frac{\chi(\hat{z}_i)}{\chi'}} = 0\;,
}
meaning that if the background lensing efficiency $g^{(j)}(\chi)$ in (\ref{eq:limber}) is replaced by (\ref{eq:modlenseff}), the contribution of matter at $\chi(\hat{z}_i)$ to the lensing signal of the background population $j$ is nulled, as desired. 

Equation (\ref{eq:zwangsbed}) only ensures that the contribution to the lensing signal is eliminated exactly at $\chi(\hat{z}_i)$, but since the lensing efficiency is a smooth function of $\chi$, the contributions from neighboring distances will also be largely downweighted. Therefore, one does not expect a perfect removal, but a substantial suppression of the GI signal due to nulling, provided that the distance probability distribution is sufficiently compact. In the still unconstrained range $0 \leq \chi \leq \chi(\hat{z}_i)$, $B^{(i)}(\chi)$ is set to zero. Henceforth, we denote the distribution in which the signal is nulled, or equivalently, the photometric redshift bin this distribution corresponds to, by \lq initial bin\rq.

Assuming disjoint, narrow bins in redshift also for (\ref{eq:limber}) by inserting (\ref{eq:bindelta}), one can define a tomography power spectrum, evaluated at precisely known comoving distances,
\eqa{
\label{eq:limberdistexact}
P_{\rm GG}(\ell;\chi_i,\chi_j) &=& \frac{9H_0^4 \Omega_{\rm m}^2}{4 c^4} \int^{\chi_{\rm hor}}_{{\rm max}(\,\chi_i,\,\chi_j)} \!\!\!\!\!\!\!\! \dd \chi\; \br{1-\frac{\chi}{\chi_i}}\; \br{1-\frac{\chi}{\chi_j}}\\ \nn
&& \hspace*{2cm} \times\; \bc{1+z(\chi)}^2 P_{\delta\delta} \br{\frac{\ell}{\chi},\chi}\;.
}
According to the modification of the lensing efficiency (\ref{eq:modlenseff}), JS08 have introduced new power spectra of the form
\eqa{
\label{eq:defPi}
\Pi^{(i)}(\ell) &=& \int_0^{\chi_{\rm hor}} \dd \chi' ~B^{(i)}(\chi')\; P_{\rm GG}(\ell;\chi(\hat{z}_i),\chi')\\ \nn
&\approx& \sum_{j=i+1}^{N_z} B^{(i)}(\chi(z_j)) ~P_{\rm GG}^{(ij)}(\ell) ~\chi'(z_j) ~\Delta z_j\;,
}
where $\Delta z_j$ denotes the width of photometric redshift bins, and where $\chi'(z)$ is the derivative of comoving distance with respect to redshift, which can be obtained analytically from (\ref{eq:wz}). The second term in (\ref{eq:defPi}) is the approximation of the foregoing integral by a Riemannian sum. It reflects the fact that information about the radial distance is available only in discrete, binned form, and in terms of redshift rather than comoving distance. Since the weight function $B^{(i)}(\chi)$ vanishes for $\chi \leq \chi(\hat{z}_i)$, the sum starts only at bin $i+1$. We will use the discrete expression of (\ref{eq:defPi}) throughout this work, including cases in which the photometric redshift bins are broad and overlapping. Transforming the constraint equation (\ref{eq:zwangsbed}) to an integral over redshift, and discretizing analogously to (\ref{eq:defPi}), one arrives at
\eq{
\label{eq:zwangsbed_discrete}
\sum_{j=i+1}^{N_z} B^{(i)}(\chi(z_j)) ~\chi'(z_j) ~\Delta z_j ~\br{1 - \frac{\chi(\hat{z}_i)}{\chi(z_j)}} = 0\;.
}

With these equations at hand we are able to demonstrate how this technique removes the GI signal. In practice, the power spectra $\Pi^{(i)}(\ell)$ will not only be composed of the lensing power spectra as written in (\ref{eq:defPi}), but of the observed signal $P_{\rm tot}^{(ij)}(\ell) = P_{\rm GG}^{(ij)}(\ell) + P_{\rm GI}^{(ij)}(\ell)$, where the latter term is unknown. Using (\ref{eq:bindelta}) again, (\ref{eq:limberGI}) is modified as follows,
\eqa{
\label{eq:PGI_approx}
P_{\rm GI}^{(ij)}(\ell) &\approx& \frac{3H_0^2 \Omega_{\rm m}}{2 c^2} ~g^{(j)}(\chi(\hat{z}_i))\; \frac{1+\hat{z}_i}{\chi(\hat{z}_i)}\; P_{\delta \gamma^{\rm I}} \br{\frac{\ell}{\chi(\hat{z}_i)},\chi(\hat{z}_i)}\\ \nn
&\approx& \frac{3H_0^2 \Omega_{\rm m}}{2 c^2}  ~\br{1 - \frac{\chi(\hat{z}_i)}{\chi(z_j)}}\; \frac{1+\hat{z}_i}{\chi(\hat{z}_i)}\; P_{\delta \gamma^{\rm I}} \br{\frac{\ell}{\chi(\hat{z}_i)},\chi(\hat{z}_i)}\;,
}
where the approximation has been applied to distribution $i$ in the first step and to distribution $j$ in the second equality. The latter transformation only affects the lensing efficiency and is readily seen by inserting the approximated distance distribution into (\ref{eq:lenseff}). Note that the second term in (\ref{eq:limberGI}), containing $g^{(i)}(\chi)~p^{(j)}(\chi)$, vanishes if the redshift distributions do not overlap. This does not hold anymore for more realistic, broader distributions, the consequences being discussed in Sect.$\,$\ref{sec:controladjacent}. Now assume that $P_{\rm GI}^{(ij)}(\ell)$, in the form as given in the second equality of (\ref{eq:PGI_approx}), adds to the lensing signal. Computing the nulled power spectrum $\Pi^{(i)}(\ell)$ according to the discrete form of (\ref{eq:defPi}), one readily finds that this new power spectrum does not have a GI contamination anymore if (\ref{eq:zwangsbed_discrete}) is fulfilled.

For the sake of a compact notation we define the vectors
\eqa{
\label{eq:defT}
\vek{T}^{(i)}_{[0]} &\equiv& \frac{\vek{T'}^{(i)}_{[0]}}{|\vek{T'}^{(i)}_{[0]}|}  ~~~\mbox{with}~~~ {{T'}_{[0]}^{(i)}}_j = \br{1 - \frac{\chi(\hat{z}_i)}{\chi(z_j)}}\;;\\ \nn
\vek{T}^{(i)}_{[1]} &\equiv& \frac{\vek{T'}^{(i)}_{[1]}}{|\vek{T'}^{(i)}_{[1]}|}  ~~~\mbox{with}~~~ {{T'}_{[1]}^{(i)}}_j = B^{(i)}(\chi(z_j)) ~\chi'(z_j) ~\Delta z_j\;,
}
so that the constraint (\ref{eq:zwangsbed_discrete}) turns into an orthogonality relation, $\br{\vek{T}^{(i)}_{[0]} \cdot \vek{T}^{(i)}_{[1]}}=0$. We now compute more weights $\vek{T}^{(i)}_{[q]}$ with $q \geq 2$ in order to construct further new power spectra of \lq order\rq\ $q$,
\eq{
\label{eq:PiwithT}
\Pi^{(i)}_{[q]}(\ell) = \sum_{j=i+1}^{N_z} {{T'}_{[q]}^{(i)}}_j ~P_{\rm tot}^{(ij)}(\ell)\;,
}
where the weights are specified by the requirement
\eq{
\label{eq:require_higherorder}
\br{\vek{T}^{(i)}_{[q]} \cdot \vek{T}^{(i)}_{[r]}} = 0  ~~~\mbox{for all}~~~ 0 \leq r < q\;.
}
In the discretized version given by (\ref{eq:zwangsbed_discrete}) the weight function has $N_z-i$ free parameters, namely the function values $B^{(i)}(\chi(z_j))$. For fixed initial bin $i$ these free parameters translate into the $N_z-i$-dimensional vectors $\vek{T}^{(i)}_{[q]}$. Since (\ref{eq:zwangsbed_discrete}) does not restrict the overall amplitude, we fix the normalization by assigning unit length to the vectors $\vek{T}^{(i)}_{[q]}$. In total, one can thus construct $N_z-i$ new power spectra per bin $i$, but since the additional constraint (\ref{eq:zwangsbed_discrete}) reduces the degrees of freedom by one, one new power spectrum cannot be freed from the GI contamination. It is the zeroth-order power spectrum, also constructed via (\ref{eq:PiwithT}) for $q=0$, which obviously cannot fulfill the nulling constraint. 

By defining vectors that contain the cosmic shear observables, i.e. in our case the power spectra,
\eqa{
\label{eq:datavectors}
\vek{P}^{(i)}(\ell) &\equiv& \bc{P_{\rm tot}^{(i,j=i+1)}(\ell), ~...~ ,P_{\rm tot}^{(i,j=N_z)}(\ell)}^\tau\;;\\ \nn
\vek{\Pi}^{(i)}(\ell) &\equiv& \bc{\Pi^{(i)}_{[0]}(\ell), ~~~~...~~~~ ,\Pi^{(i)}_{[N_z-i-1]}(\ell)}^\tau
}
and composing the transformation matrix
\eq{
\label{eq:transformationmatrix}
\vek{T}^{(i)} \equiv \br{\vek{T}^{(i)}_{[0]}, ... ,\vek{T}^{(i)}_{[N_z-i-1]}}
}
for every distribution $i$ and angular frequency $\ell$, the new power spectra are given by $\vek{\Pi}^{(i)}(\ell) = \vek{T}^{(i)} \vek{P}^{(i)}(\ell)$. Due to the construction of the weights $\vek{T}^{(i)}_{[q]}$ the transformation matrix is orthogonal, and so is the transformation of the full data set. Therefore the nulling technique can be interpreted as a rotation of the cosmic shear data vector such that in the rotated set the GI contamination is restricted to certain elements, namely those with a subscript $[0]$. By removing these, one loses part of the lensing signal and hence statistical power, but eliminates the GI systematic within the limits of the approximations made in the foregoing derivation.

Performing a rotation, the dimension of the nulled data vector, which is composed of the $\vek{\Pi}^{(i)}(\ell)$ for every $i$ and $\ell$, is exactly the same as for the original data set. For the data analysis one removes the contaminated nulled power spectra with subscript $\bb{0}$, i.e. one entry per initial bin. This is the step that actually does the nulling and modifies both statistical and systematic error budgets. In this work, we are going to use all remaining nulled power spectra with $q \geq 1$ throughout. Since they are merely specified by being composed of mutually orthogonal weights, there is no ordering among different $q$. In particular, it is impossible to make a priori statements about the information content of different orders $q$.

It should be noted, however, that one can combine the formalism outlined above with a data compression algorithm, based on Fisher information. As investigated in JS08, nearly all information about cosmological parameters can be concentrated in a limited set of nulled power spectra, constructed from the first-order weights $\vek{T}^{(i)}_{[1]}$. The additional requirement that a suitable combination of Fisher matrix elements is to be maximized introduces a strong hierarchy in terms of information content into the sequence of $\vek{\Pi}^{(i)}(\ell)$ with $q \geq 1$. We will not consider such an optimization in this work.

\subsection{Fisher matrix formalism}
\label{sec:fisher}

In the following analysis we will make use of the Fisher matrix formalism (see \citealp{tegmark97} for details) to obtain parameter constraints. Probing the likelihood locally around its maximum, it is computationally much cheaper than a full likelihood analysis and thus useful for error estimates for a large set of models. The elements of the Fisher matrix are defined by
\eq{
\label{eq:fisherdef}
F_{\mu\nu} = - \ba{ \frac{\partial^2 \ln L}{\partial p_\mu ~\partial p_\nu} }\;,
}
for a set of parameters $\vek{p}$, where $L$ denotes the likelihood. In this paper the set of cosmological parameters $\bc{\Omega_{\rm m},\sigma_8,h_{100},n_{\rm s},\Omega_{\rm b},w_0,w_a}$ is considered, see Sect.$\,$\ref{sec:ps} for further details.

To second-order Taylor expansion around the maximum likelihood point the likelihood can be described by a multivariate Gaussian, so that, as long as only regions in parameter space are probed where the non-Gaussian contributions are negligible, it is sufficient to consider a Gaussian likelihood
\eqa{
\nn
L_x(\vek{x}|\vek{p}) &=& \frac{1}{(2\pi)^\frac{N_d}{2} \sqrt{\det C_x(\vek{p})}}\\
\label{eq:gaussianlike}
&& \times\; \exp \bc{-\frac{1}{2} \bb{\vek{x}-\vek{\bar{x}}(\vek{p})}^\tau {C_x(\vek{p})}^{-1} \bb{\vek{x}-\vek{\bar{x}}(\vek{p})}}
}
for a data vector $\vek{x}$ with expectation value $\vek{\bar{x}}(\vek{p})$ and covariance $C_x(\vek{p})$, where $N_d$ is the dimension of the full data vector. \citet{tegmark97} have shown that for this case the Fisher matrix reads
\eqa{
\nn
F_{\mu\nu} &=& \frac{1}{2}\; {\rm tr} \Biggl\{ {C_x}^{-1}\, \frac{\partial C_x}{\partial \mu}\, {C_x}^{-1}\, \frac{\partial C_x}{\partial \nu}\\ 
\label{eq:gaussianfisher}
&& \hspace*{2.5cm} + {C_x}^{-1} \br{ \frac{\partial \vek{\bar{x}}}{\partial \mu}\, \frac{\partial \vek{\bar{x}}^\tau}{\partial \nu} + \frac{\partial \vek{\bar{x}}}{\partial \nu}\, \frac{\partial \vek{\bar{x}}^\tau}{\partial \mu}} \Biggr\}\;,
}
where the argument of $\vek{\bar{x}}$ and $C_x$ has been omitted for convenience.

Now consider an invertible linear transformation $\vek{T}$ of the data vector,
\eq{
\label{eq:datatrafo}
\vek{y} \equiv \vek{T} \vek{x}\;; \hspace*{.5cm} C_y = \vek{T} C_x \vek{T}^\tau\;.
}
In this work, $\vek{x}$ corresponds to the data vector $\vek{P}^{(i)}(\ell)$, and $\vek{y}$ to the nulled data vector $\vek{\Pi}^{(i)}(\ell)$, while the transformation is given by (\ref{eq:PiwithT}). Plugging the relations (\ref{eq:datatrafo}) into (\ref{eq:gaussianlike}), one finds that the exponential remains unchanged, while the prefactor gets an additional term $|\det \vek{T}|^{-1}$, using $\det \br{\vek{T} C_x \vek{T}^\tau} = \det C_x \det^2 \vek{T}$. This modification merely leads to a rescaling of the likelihood values, and thus likelihood contours in parameter space remain unchanged. Since $\vek{T}$ is invertible, the data in $\vek{x}$ and $\vek{y}$ contains the same amount of information about the parameters. Accordingly, the Fisher matrix is also invariant under this transformation \citep{tegmark97}, which is easily demonstrated by inserting (\ref{eq:datatrafo}) into (\ref{eq:gaussianfisher}).

However, in the case of nulling the transformation (\ref{eq:PiwithT}) to the new data vector $\vek{\Pi}^{(i)}(\ell)$ depends on the cosmological parameters one aims at determining because the elements of $\vek{T}$ are composed of comoving distances. Hence, the likelihood is now parameter-dependent in both arguments,
\eq{
\label{eq:likey}
L_y(\vek{y}|\vek{p}) = \br{\det \vek{T}(\vek{p})}^{-1}\; L_x(\vek{x}|\vek{p})\;,
}
where we omitted the modulus of $\det \vek{T}$ as this expression can always be turned positive by swapping two entries of either the original or the transformed data vector. The prefactor in (\ref{eq:likey}) acts like a prior on the original likelihood of $\vek{x}$. In JS08 an example of the magnitude of the effect of this prior was assessed unintentionally by not taking into account the prefactor although $\det \vek{T}$ differed from unity due to a different normalization. As stated in JS08, however, the likelihood values of both data sets were checked to be identical to the level of numerical accuracy. We conclude that the effect of the prior due to the data transformation must have been considerably weaker than the one of the flat prior imposed in the analysis. As far as nulling is concerned, the prior of (\ref{eq:likey}) only acts on cosmological parameters that enter (\ref{eq:wz}) in a non-trivial way.

We intend to compute the Fisher matrix for the original and the transformed data set, in both cases at the point of maximum likelihood, i.e. for the fiducial set of parameters. At this point in parameter space we expect the derivative with respect to parameters to vanish on average, $\ba{\partial L/\partial p_\mu}=0$. If the relation holds for $L_x(\vek{x}|\vek{p})$, it is clear from (\ref{eq:likey}) that this is generally not the case for $L_y(\vek{y}|\vek{p})$. Therefore we set the requirement that $\det \vek{T} = 1$, which is fulfilled by the orthogonal transformation constructed in the foregoing section. Then one can show that the Fisher matrices of both data vectors are equivalent, even for a parameter-dependent data transformation, as is detailed in App.$\,$\ref{sec:appendix}.

Furthermore, we assume that the original covariance $C_x$ does not depend on cosmological parameters. Since an additional cosmology dependence would lead to tighter constraints, this is a conservative assumption \citep[see e.g.][]{eifler08}. Using the equivalence of the Fisher matrices, and returning to the notation in the context of the nulling technique, we then arrive from (\ref{eq:gaussianfisher}) at the following expression for the original (index \lq orig\rq) and the nulled (index \lq null\rq) data vector (see App.$\,$\ref{sec:appendix}),
\eqa{
\label{eq:fisherfinal}
F^{\rm orig}_{\mu \nu} &=& \sum_{\alpha,\,\beta=1}^{N_{\rm d}} \frac{\partial {P_{\rm GG}}_\alpha}{\partial p_\mu} \br{C_P^{-1}}_{\alpha \beta} \frac{\partial {P_{\rm GG}}_\beta}{\partial p_\nu} \\ \nn
 &=& \sum_{\alpha,\,\beta,\, \gamma,\, \delta=1}^{N_{\rm d}} T_{\alpha\gamma}\; \frac{\partial {P_{\rm GG}}_\gamma}{\partial p_\mu} \br{C_\Pi^{-1}}_{\alpha \beta} T_{\beta\delta}\; \frac{\partial {P_{\rm GG}}_\delta}{\partial p_\nu} \equiv F^{\rm null}_{\mu \nu}\;,
}
where $\vek{P_{\rm GG}}$ and $\vek{T}$ are the lensing power spectrum data vector and the nulling transformation matrix of the full data set, respectively. The data vectors of the full set have the dimension $N_{\rm d}=N_\ell N_z \br{N_z-1}/2$ if $N_\ell$ angular frequency bins are considered. The covariance matrices of the original and nulled power spectra are denoted by $C_P$ and $C_\Pi$. The equality of original and nulled Fisher matrix, i.e. the Fisher matrix after performing the nulling rotation, directly follows from (\ref{eq:datatrafo}), second equation. However, the actual nulling step removes elements from the transformed data vector, thereby reducing the dimension of the nulled data vector to $N_\ell \br{N_z-1} \br{N_z-2}/2$ and causing $F^{\rm null,red}_{\mu \nu} \leq F^{\rm orig}_{\mu \nu}$, where $F^{\rm null,red}_{\mu \nu}$ denotes the Fisher matrix, computed from the nulled data vector after the removal of the contaminated power spectra with $q=0$.

Since the inverse Fisher matrix is an estimate for the parameter covariance matrix, we compute the marginalized statistical errors as $\sigma(p_\mu) = \sqrt{(F^{-1})_{\mu\mu}}$. Due to the Cram\'er-Rao inequality this is a lower bound on the error. To assess the effect of the systematic, we also calculate the bias on every parameter by means of the bias formalism \citep{kim04,huterer05b,huterer06,taylor07,amara08,kitching08}. Assuming a systematic $P_{\rm GI}$ that is subdominant with respect to the signal and causes only small systematic errors, the bias $b$ on a parameter $p_\mu$ can be calculated by
\eq{
\label{eq:bias}
b(p_\mu) = \sum_\nu \br{ F^{\rm orig}_{\mu \nu} }^{-1} \sum_{\alpha,\,\beta=1}^{N_{\rm d}} {P_{\rm GI}}_\alpha \br{C_P^{-1}}_{\alpha \beta} \frac{\partial {P_{\rm GG}}_\beta}{\partial p_\nu}\;,
} 
and likewise for the nulled data set. A formal derivation of the bias formalism, including the discussion of its limitations can be found in App.$\,$\ref{sec:validitybias}.

\section{Modeling}
\label{sec:modelling}

\subsection{Redshift distributions}

To model realistic redshift probability distributions of galaxies in the presence of photometric redshift errors, we keep close to the formalisms used in \citet{ma05} and \citet{amara07}. We assume survey parameters that should be representative of any future space-based mission aimed at precision measurements of cosmic shear, such as the Euclid satellite proposed to ESA. Note that the probability distributions of comoving distances and redshift, used in parallel in this work, are related via $p_z(z)=p_\chi(\chi)\,\chi'(z)$. 

According to \citet{smail94} we assume an overall redshift probability distribution 
\eq{
\label{eq:redshiftdistribution}
p_{\rm tot}(z) \propto \br{\frac{z}{z_0}}^2 \exp \bc{ -\br{\frac{z}{z_0}}^\beta}
}
with $\beta=1.5$. To get a median redshift of $z_{\rm med}=0.9$, we choose $z_0=0.64$. The distribution is cut at $z_{\rm max}=3$ and then normalized to unity. The total distribution of galaxies per unit survey area is then $n_{\rm tot}(z)=n\, p_{\rm tot}(z)$, where $n$ is the total number density of galaxies. The choice of photometric redshift bin boundaries for the tomography is in principle arbitrary. Here, we divide $p_{\rm tot}(z)$ into $N_z$ photometric redshift bins such that every bin contains the same number of galaxies, i.e. 
\eq{
\label{eq:dividebins}
\int_{z_{i-1}}^{z_i} \dd z\; p_{\rm tot}(z) = \frac{1}{N_z} ~~~\mbox{for every}~~ i=1,\,...\,,N_z\;, 
}
where the $z_i$ mark the redshifts of the bin boundaries, and where $z_0=0$ and $z_{N_z}=z_{\rm max}$. This choice of binning is solely for computational convenience and to allow for easy comparisons of setups with a different number of bins. The nulling technique as such does not rely on any particular choice of photometric redshift binning.

\begin{figure}[ht]
\centering
\includegraphics[scale=.6]{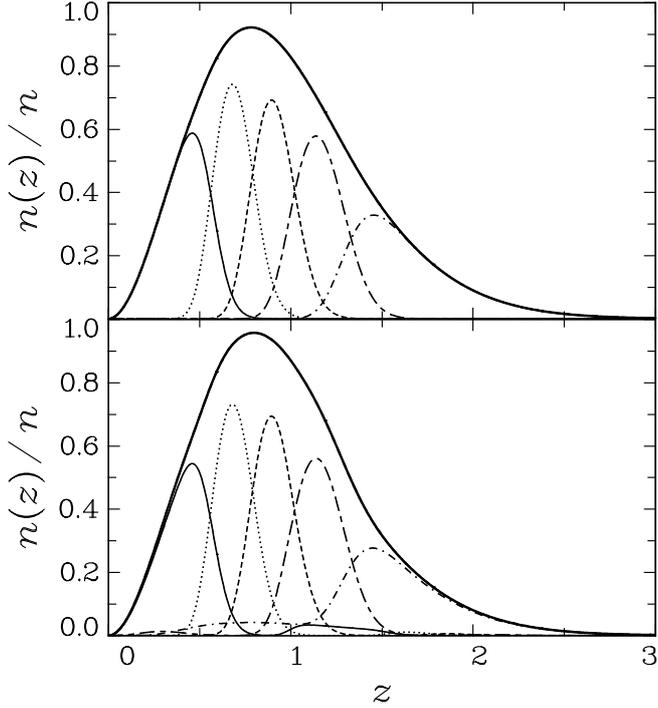}
\caption{Number density distribution of galaxies for a division into $N_z=5$ redshift bins, rendered dimensionless through dividing by the total number density $n$. The thick solid line corresponds to the overall galaxy number density distribution, normalized to unity. The thin curves represent the distributions corresponding to the five photometric redshift bins, normalized to $1/N_z$. The original bin boundaries are chosen according to (\ref{eq:dividebins}). Note that the sum of the individual distributions adds up to the total distribution for every $z$. \textit{Top panel}: Resulting distributions for $\sigma_{\rm ph}=0.05$ and no catastrophic outliers. \textit{Bottom panel}: Resulting distributions for $\sigma_{\rm ph}=0.05$, $f_{\rm cat}=0.1$, and $\Delta_z=1.0$.}
\label{fig:zdistributions}
\end{figure}

Our model for photometric redshift errors accounts for two effects, a statistical uncertainty characterized by the redshift dispersion $\sigma_{\rm ph}(1+z)$, and misidentifications of a fraction $f_{\rm cat}$ of galaxies with offsets from the center of the distribution of $\pm \Delta_z$. We write the conditional probability of obtaining a photometric redshift $z_{\rm ph}$ given the true, spectroscopic redshift $z$ as
\eqa{
\nn
p(z_{\rm ph}\,|\,z) &\propto& \br{1-f_{\rm cat}}\; G\br{z_{\rm ph};\;z,\,\sigma_{\rm ph}\br{1+z}} + \frac{f_{\rm cat}}{2}\\
\label{eq:zphgivenzspec} 
&& \hspace{-1.5cm} \times\; \bc{G\br{z_{\rm ph};\;z_+,\,\sigma_{\rm ph}\br{1+z_+}} + G\br{z_{\rm ph};\;z_-,\,\sigma_{\rm ph}\br{1+z_-}}}\;,
}
where $G\br{z_{\rm ph};\;z,\,\sigma}$ is a Gaussian with mean $z$ and dispersion $\sigma$, and where $z_+=z+\Delta_z$ and $z_-=z-\Delta_z$. When integrating (\ref{eq:zphgivenzspec}) over $z_{\rm ph}$ with infinite range, it yields unity for every $z$. However, since we consider a finite redshift range, the distributions corresponding to the lowest and highest photometric redshift bins and those with significant outlier population will be cut at 0 and $z_{\rm max}$, so that we normalize $p(z_{\rm ph}\,|\,z)$ by demanding $\int_0^{z_{\rm max}} \dd z_{\rm ph}\; p(z_{\rm ph}\,|\,z) = 1$ for every $z$. Multiplying $p(z_{\rm ph}\,|\,z)$ with the overall redshift probability distribution of galaxies $p_{\rm tot}(z)$ yields the two-dimensional probability of obtaining a pair of redshift measurements $\{z_{\rm ph},z\}$. When integrating this probability over photometric redshift within the bin boundaries defined above, one arrives at the true probability distribution of galaxies for every photometric redshift bin $i$,
\eq{
\label{eq:binzdis}
p^{(i)}(z) = \frac{ p_{\rm tot}(z) \int_{z_{i-1}}^{z_i} \dd z_{\rm ph}\; p(z_{\rm ph}\,|\,z) }{ \int_0^{z_{\rm max}} \dd z'\; p_{\rm tot}(z') \int_{z_{i-1}}^{z_i} \dd z_{\rm ph}\; p(z_{\rm ph}\,|\,z')  }\;.
}
Due to the multiplication by $p_{\rm tot}(z)$ these distributions are limited to the interval $\bb{0,z_{\rm max}}$ although (\ref{eq:zphgivenzspec}) is non-vanishing outside that range. To ensure that the dispersions of the Gaussians in (\ref{eq:zphgivenzspec}) are positive, $\Delta_z \leq 1$ is required. In this work we set $\Delta_z=1$ fixed since this choice produces outlier distributions that are well separated from the central peak, as also found in realistic situations, see below.

The number density of galaxies located in photometric redshift bin $i$ as a function of spectroscopic redshift is given by
\eq{
\label{eq:binndis}
n^{(i)}(z) = n_{\rm tot}(z) \int_{z_{i-1}}^{z_i} \dd z_{\rm ph}\; p(z_{\rm ph}\,|\,z)\;,
}
so that evidently $\sum_i n^{(i)}(z) = n_{\rm tot}(z)$ for every redshift $z$. Using this last equation and multiplying (\ref{eq:dividebins}) by $n$, one sees that the sum of the number densities of galaxies, having their true redshifts between the bin boundaries defined by (\ref{eq:dividebins}), is the same for all bins, namely $n/N_z$, as requested. However, the number densities of galaxies per photometric redshift bin, i.e. $n^{(i)} = \int_0^{z_{\rm max}} \dd z\; n^{(i)}(z)$, are generally not identical. The photometric redshift errors lead to a redistribution of galaxies, which will in our model cause the outermost galaxy distributions to contain slightly more objects than $n/N_z$.

Two examples for galaxy distributions $n^{(i)}(z)$ obtained via this formalism are shown in Fig.$\,$\ref{fig:zdistributions}, one without outliers and with a dispersion of $\sigma_{\rm ph}=0.05$, and one where outliers with $f_{\rm cat}=0.1$ at an offset $\Delta_z=1$ have been added. As is evident from the plot in the lower panel, the outlier Gaussians are modified by (\ref{eq:binzdis}) into elongated bumps, which are well separated from the central peak. They are most prominent as a distribution with $z \gtrsim 1$, being part of the lowest photometric bin, and a broad distribution at low redshifts, belonging to the highest photometric bin. This behavior is qualitatively in good agreement with the characteristic shape of the scatter plots in the spectroscopic redshift - photometric redshift plane, as for instance analyzed in \citet{abdalla07}, which also justifies our choice of $\Delta_z=1$. 

\begin{figure}[t]
\centering
\includegraphics[scale=.6,angle=270]{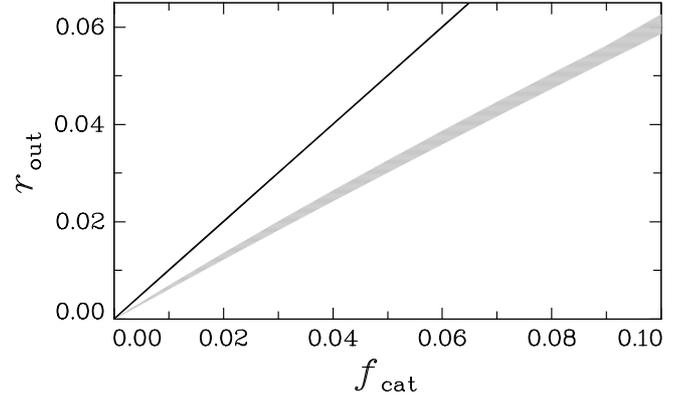}
\caption{Relation between $f_{\rm cat}$ and the true fraction of outliers in the redshift distributions $r_{\rm out}$. The gray area marks the range of possible values of $r_{\rm out}$ if $\sigma_{\rm ph}$ lies in the interval $\bb{0.01;\,0.1}$, where $\sigma_{\rm ph}=0.01$ produces the upper limit and $\sigma_{\rm ph}=0.1$ the lower limit of the gray region. A one-to-one relation is indicated by the solid black line.}
\label{fig:realfcat}
\end{figure}

To judge the performance of nulling in the presence of catastrophic outliers in the redshift distributions, it is important to note that $f_{\rm cat}$ does not equal the true fraction of outliers, primarily because of the subsequent multiplication of (\ref{eq:zphgivenzspec}) by the overall redshift distribution $p_{\rm tot}(z)$, see (\ref{eq:binzdis}). We compute the true fraction of outliers, denoted by $r_{\rm out}$, as the part of a redshift distribution that is contained in the two outlier Gaussians of our model. A quantity $p_{\rm cat}(z_{\rm ph}\,|\,z)$ is defined identically to (\ref{eq:zphgivenzspec}), but with the first term, i.e. the central Gaussian, removed. Then we define the outlier fraction as
\eq{
\label{eq:defrout}
r_{\rm out} \equiv \frac{1}{N_z} \sum_{i=1}^{N_z} \frac{ \int_0^{z_{\rm max}} \dd z\; p_{\rm tot}(z) \int_{z_{i-1}}^{z_i} \dd z_{\rm ph}\; p_{\rm cat}(z_{\rm ph}\,|\,z) }{ \int_0^{z_{\rm max}} \dd z\; p_{\rm tot}(z) \int_{z_{i-1}}^{z_i} \dd z_{\rm ph}\; p(z_{\rm ph}\,|\,z) }\;,
}
where $r_{\rm out}$ is averaged over all photometric redshift bins.

In Fig.$\,$\ref{fig:realfcat} the relation between $r_{\rm out}$ and $f_{\rm cat}$ for fixed $\Delta_z=1.0$ is plotted. The gray region comprises the results for the range from $\sigma_{\rm ph}=0.01$ to $\sigma_{\rm ph}=0.1$. Evidently, the true fraction of outliers is smaller than $f_{\rm cat}$, reaching up to about $6\,\%$ for $f_{\rm cat} \leq 0.1$. The strongest contribution to $r_{\rm out}$ originates from the bins at the lowest and highest redshifts, where the outlier distributions are enhanced because one of the outlier Gaussians is located in a redshift regime where $p_{\rm tot}(z)$ obtains high values. The redshift distributions centered at medium redshifts have their central Gaussian at $z \sim 1$ where $p_{\rm tot}(z)$ peaks, so that the outlier fraction in the corresponding bins is small. 

In the following, we will consider the range $0 \leq f_{\rm cat} \leq 0.1$, which yields outlier fractions that should comprise realistic limits of catastrophic failures in the photometric redshift determination of surveys aimed at measuring cosmic shear tomography \citep[see][]{abdalla07}. For the COSMOS field \citet{ilbert08} found photometric redshift dispersions in the range between 0.007 for the brightest galaxies and 0.06 for fainter objects up $z\sim2$. Taking these values as a reference, we are going to consider the range $0 \leq \sigma_{\rm ph} \leq 0.1$.

\subsection{Lensing power spectra}
\label{sec:ps}

As the basis for our analysis we use sets of tomography lensing power spectra which are computed for a $\Lambda$CDM universe with fiducial parameters $\Omega_{\rm m}=0.25$, $\Omega_{{\rm DE},0}=0.75$, and $H_0=100\, h_{100}\, {\rm km/s/Mpc}$ with $h_{100}=0.7$. Throughout, the spatial geometry of the Universe is assumed to be flat. We incorporate a variable dark energy scenario by parametrizing its equation of state, relating pressure $p_{\rm DE}$ to density $\rho_{\rm DE}$, as
\eq{
\label{DEeos}
p_{\rm DE} = \br{w_0 + w_a \frac{z}{1+z}}\; \rho_{\rm DE} c^2\;,
}
where the cosmological constant is chosen as the fiducial model, i.e. $w_0=-1$ and $w_a=0$. Then the dark energy density parameter reads
\eq{
\label{eq:darkenergy}
\Omega_{\rm DE}(z) = \Omega_{{\rm DE},0}\; \exp 3 \br{w_a \frac{z}{1+z} - \br{w_0 + w_a + 1} \ln (1+z)}\;.
}
The three-dimensional power spectrum of matter density fluctuations $P_{\delta\delta}$ is further specified by the primordial slope $n_{\rm s}=1$, the normalization $\sigma_8=0.9$ and the shape parameter $\Gamma$, calculated according to \citet{sugiyama95} with $\Omega_{\rm b}=0.05$. Using the transfer function of \citet{eisenstein98} (without baryonic wiggles), the non-linear power spectrum is computed by means of the fit formula of \citet{PeacockDodds}. The tomography power spectra are then determined via (\ref{eq:limber}), incorporating the photometric redshift models of the foregoing section, for $N_\ell=100$ logarithmic angular frequency bins between $\ell=10$ and $\ell=2\cdot10^4$.

\begin{figure*}[t]
\centering
\includegraphics[scale=.77,angle=270]{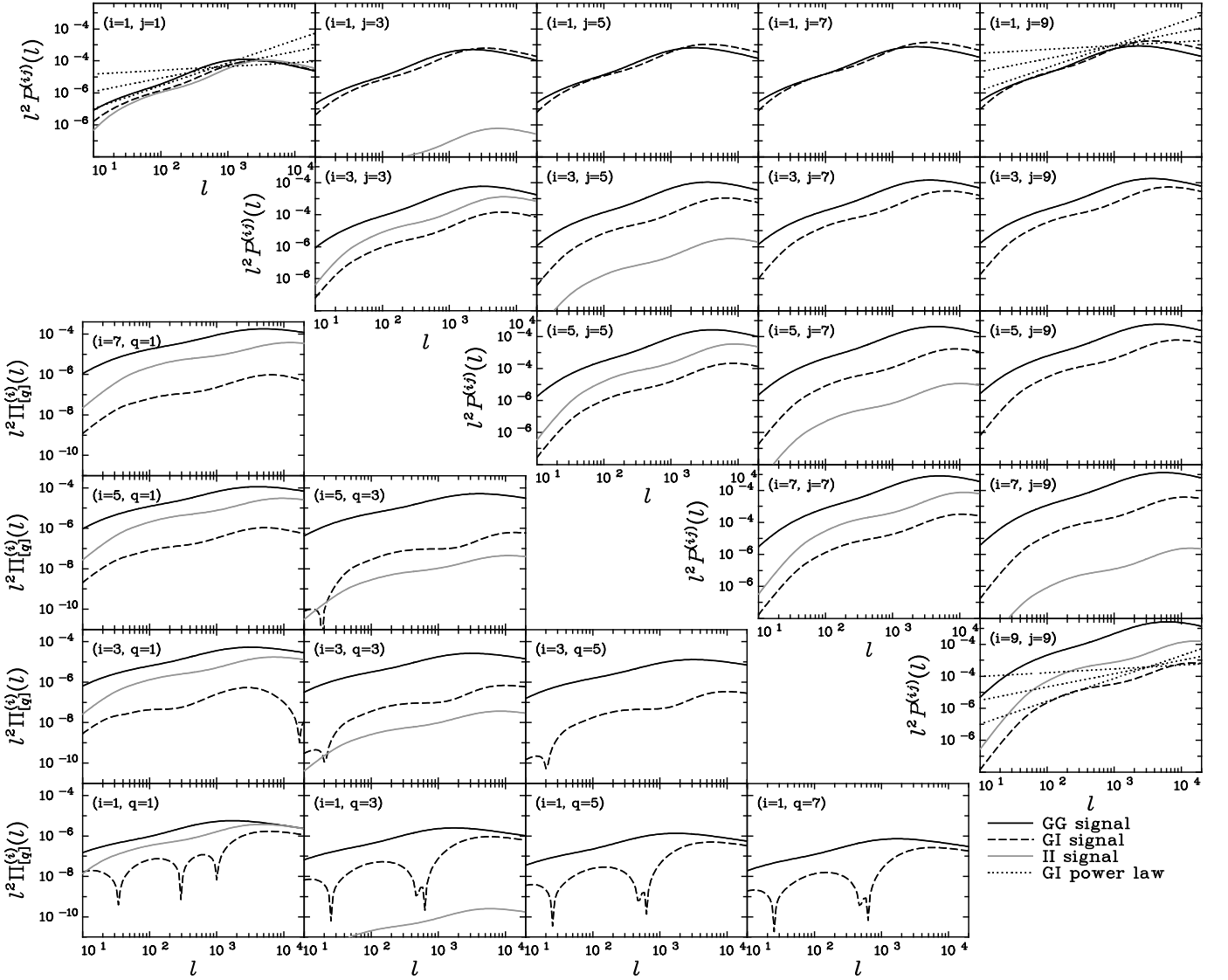}
\caption{Original and nulled tomography power spectra as a function of angular frequency. The survey has been divided into $N_z=10$ photometric redshift bins with dispersion $0.03(1+z)$. \textit{Top right panels}: Lensing power spectra $P^{(ij)}_{\rm GG}(\ell)$ are shown as solid lines. The modulus of linear alignment model GI power spectra $P^{(ij)}_{\rm GI}(\ell)$ is given by dashed lines, the corresponding II signal by gray curves. In each panel the redshift bins $i$ and $j$ are plotted. In the panels with the combinations $i,j \in \bc{1,9}$ the absolute values of the power law GI models have been added for reference as dotted curves. Note that the II power spectrum becomes very small if $i$ and $j$ are largely different. \textit{Bottom left panels}: The absolute values of the nulled lensing and linear alignment model systematic power spectra are shown as solid (GG), dashed (GI), and gray (II) curves, respectively. In each panel the corresponding redshift bin $i$ and the order $q$ are given. The nulled measures do not have a particular ordering in $q$, see text for details. For the lower redshift bins the GI signal is oscillating around zero. The II signal becomes very small for higher orders $q$.}
\label{fig:ps}
\end{figure*}

The nulled power spectra $\Pi^{(i)}_{[q]}(\ell)$ are then calculated via (\ref{eq:PiwithT}). The nulling weights $\vek{T}^{(i)}_{[0]}$, see (\ref{eq:defT}), are computed for the fiducial cosmology, while the higher orders are obtained by Gram-Schmidt ortho-normalization. The Gram-Schmidt procedure does not uniquely define the order of the orthogonal vectors, so that no particular ordering is assigned to $q$, as opposed to the approach in JS08, where a higher order $q$ corresponded to a lower information content in $\Pi^{(i)}_{[q]}(\ell)$.

On applying nulling to a real data set, one has to assume the values of the relevant parameters $\Omega_{\rm m}$, $\Omega_{\rm DE}$, $w_0$, and $w_a$ to obtain $\vek{T}^{(i)}_{[0]}$. Whilst it is a realistic premise that these parameters are approximately known, slightly incorrect assumptions may degrade the downweighting of the GI signal, but do not introduce a new bias to the parameter estimation, as will be assessed in detail in Sect.$\,$\ref{sec:cosdep}. A sample of both original and nulled tomography power spectra are plotted in Fig.$\,$\ref{fig:ps}. For this sample the nulling has been performed following variant (C), which will be discussed in detail in Sect.$\,$\ref{sec:optnulling}.

As regards the calculation of the power spectrum covariance \citep[and references therein]{joachimi08}, entering the Fisher matrix, we have to specify further survey characteristics in addition to the aforementioned redshift probability distribution. We assume a survey size of $20,000\,{\rm deg}^2$ and a total number density of galaxies of $n=35\,{\rm arcmin}^{-2}$, resulting in approximately $35/N_z\,{\rm arcmin}^{-2}$ galaxies per photometric redshift bin. To compute shot noise, the dispersion of intrinsic ellipticities is set to $\sigma_\epsilon=0.35$. These survey parameters correspond to those representative of future cosmic shear satellite missions such as Euclid.

\subsection{Intrinsic alignment signal}
\label{sec:GIsignal}

To quantify the bias on cosmological parameters before and after nulling, a GI systematic power spectrum is added to the data vector. We adopt the \lq non-linear linear alignment model\rq\ of \citet{bridle07}, who suggest to compute the three-dimensional matter-intrinsic shear cross-power spectrum as 
\eq{
\label{eq:GIlinalign}
P^{\rm l.a.}_{\delta \gamma^{\rm I}}\br{k,z} = - C_{\rm GI}\; \rho_{\rm cr}\; \frac{\Omega_{\rm m}\;(1+z)^2}{D(z)}\; P_{\delta \delta}\br{k,z}\;,
}
where $\rho_{\rm cr}$ is the critical density, and where $D(z)$ denotes the growth factor, normalized to unity for $z=0$. The constant $C_{\rm GI}$ has units of inverse density and was determined by HS04 through comparison with SuperCOSMOS \citep{brown02}; according to \citet{bridle07}, we set $C_{\rm GI}\, \rho_{\rm cr} \approx 0.0134$. The corresponding II power spectrum reads
\eq{
\label{eq:IIlinalign}
P^{\rm l.a.}_{\gamma^{\rm I} \gamma^{\rm I}}\br{k,z} = C^2_{\rm GI}\; \rho^2_{\rm cr}\; \frac{\Omega^2_{\rm m}\;(1+z)^4}{D^2(z)}\; P_{\delta \delta}\br{k,z}\;.
}
Originating from analytical considerations by HS04, the linear alignment model in the form employed here lacks solid physical motivation, but fits within the error bars of \citet{mandelbaum06}. It also provides reasonable fits to the results of the halo model considerations by \citet{schneiderm09}.

While the nulling technique as such is completely independent of the actual functional form of the systematic, the residual bias does depend on the GI signal. Thus, we consider an additional set of simplistic power-law GI power spectra for reference. They are given by
\eq{
\label{eq:GIpowerlaw}
P^{\rm p.l.}_{\delta \gamma^{\rm I}}\br{k,z} = - A_{\rm GI}\; \br{\frac{k}{k_{\rm ref}}}^{s-2} (1+z)\;,
}
where $k_{\rm ref}=1\,h_{100}/{\rm Mpc}$. As is evident from (\ref{eq:bias}), the produced bias is simply proportional to the amplitude of the systematic, so that we do not need to investigate variations of the overall magnitude of the GI term. Hence, we relate the normalization of (\ref{eq:GIpowerlaw}) to the linear alignment model (\ref{eq:GIlinalign}), and set $A_{\rm GI}=|P^{\rm l.a.}_{\delta \gamma^{\rm I}}(k_{\rm ref},z_{\rm med})| (1+z_{\rm med})^{-1}$. For the power law slope we use the values $s=\bc{0.1,0.4,0.7}$, where the central value best reproduces the average slope of the linear alignment model power spectra. The tomography power spectra are then obtained via (\ref{eq:limberGI}).

The resulting power spectra are also shown in Fig.$\,$\ref{fig:ps}. As already mentioned in \citet{bridle07}, the linear alignment model produces a strong systematic, partially surpassing the lensing signal in amplitude for cross-correlations of largely different redshift bins. Since the GI term is negative, the sum of lensing and intrinsic alignment power spectrum can become negative in the corresponding $\ell$-range in these cases\footnote{Note however that the total power spectrum of auto-correlations of ellipticities, i.e. GG+GI+II, always has to be positive by definition.}. Due to our choice of normalization, the power-law toy GI signal can dominate the lensing power spectrum on even larger angular frequency intervals.

After nulling, the systematic is largely suppressed, oscillating around zero for the lower redshift bins. Still, significant residual signals remain because the finite extent of the redshift probability distributions has been neglected in the derivation of nulling. In particular, the systematic signal is eliminated only at a single redshift within each bin, thus being merely downweighted in neighboring redshift ranges. A detailed discussion about the sources of the residual bias will follow in Sect.$\,$\ref{sec:zinfo}. We note that nulling works independently of the strength of the systematic; it can even be applied to data in which the GI term surpasses the cosmic shear signal.

We have also added II power spectra to Fig.$\,$\ref{fig:ps} in order to judge in how far our assumption of dropping the II signal in our considerations is valid. The original II power spectra yield a strong contribution for auto-correlations, but drop off quickly if the correlated redshift distributions have less overlap. In the transformed data set, the II contamination is smaller than the residual GI signal and thus negligible for power spectra with $q > 1$. For $q=1$ however, the II signal is significant such that in this case nulling would have to be preceded by an II removal technique. In the limit of completely disjoint photometric bins, the II signal would be confined to auto-correlations in the original data set. Since these are not included into the construction of the nulled power spectra, the latter would be completely free of II terms in this idealized case.

\begin{table}[t]
\caption{Upper limits on the allowed angular frequency range if the II contamination in the nulled data shall be suppressed by at least a factor of $s$ with respect to the nulled GG term. These limitations apply only for orders $q=1$, and only if nulling is not preceded by a suitable II removal technique, as we advocate. The parameters are the same as in Fig.$\,$\ref{fig:ps}. Note that in a narrow range around $\ell \sim 100$ the II signal can be close to or slightly above the limit imposed by $s$.}
\centering
\begin{tabular}[t]{c|cc}
initial bin $i$ & $s=3$ & $s=5$\\
\hline
\hline
1 & 1170  & 20 \\
2 & 3420  & 1470 \\
3 & 5420  & 2330 \\
4 & 7960  & 3170 \\
5 & 11680 & 4310 \\
6 & none  & 5860 \\
7 & none  & 7960 \\
8 & none  & 13620 \\
\end{tabular}
\label{tab:IIlimits}
\end{table}

To ensure that the II term remains sufficiently small compared to the GG signal, one could restrict the subsequent analysis partly to larger angular scales. For instance, to achieve a minimum suppression by a factor $s$ of the II signal with respect to the lensing signal, we determine maximum allowed $\ell$-values, given in Table \ref{tab:IIlimits}. These upper bounds would only have to be applied to orders $q=1$, and are valid in the case of the setup used to produce Fig.$\,$\ref{fig:ps}. The limitations due to the II contamination are expected to become more restrictive as the photometric redshift scatter increases. 

Alternatively, our findings suggest that, due to the confinement of the II term to a limited set of nulled power spectra, a treatment of the II signal \textit{after} nulling may also provide a promising ansatz. In the current implementation the nulled power spectra of order $q=1$ have a dominating contribution from original power spectra $P^{ij}(\ell)$ with $j=i+1$, which contain the bulk of the II signal after the removal of auto-correlations from the analysis. Hence, the residual II terms accumulate within the measures of order $q=1$. The freedom to choose the weights of (\ref{eq:PiwithT}) in the subspace orthogonal to $\vek{T}^{(i)}_{[0]}$ allows for a more specific treatment of the II signal in the nulled data. We emphasize that the final goal is a simultaneous removal of all intrinsic alignment contributions, but this is beyond the scope of this paper and subject to future work.

As the GI contamination has a large amplitude, the question is raised whether the bias formalism, i.e. (\ref{eq:bias}), still yields accurate results. The effect of a large systematic is investigated in detail in App.$\,$\ref{sec:validitybias}. We conclude from our findings that even for a strong GI term the bias is obtained with good accuracy whereas the statistical errors, which are also affected by a strong systematic, can deviate more significantly. To guarantee results that are as close as possible to a full likelihood analysis, we downscale all GI signals by a factor of five throughout the subsequent sections. Since the bias is proportional to the overall amplitude of the systematic, and since we are mostly going to consider ratios of biases, the rescaling does not have an influence on the statements concerning the performance of nulling. Merely the mean square error, defined by
\eq{
\label{eq:sigmatot}
\sigma_{\rm tot}(p_\mu)=\sqrt{\sigma^2(p_\mu)+b^2(p_\mu)}\;,
}
is affected because the systematic error becomes less dominant. A lower systematic amplitude slightly disfavors nulling as it lowers the bias while causing an increase in statistical errors. Besides, limiting the strength of biases avoids unphysical parameter estimates as for instance $\Omega_{\rm m} < 0$. Such effects are normally avoided by priors, which have not been included in our Fisher matrix analysis though.

In surveys with a significant GI systematic, intrinsic ellipticity correlations are likely to affect parameter estimation, too. To restrict our considerations to the GI contamination, we follow \citet{takada04b}, excluding auto-correlations from both original and nulled data vectors, and assuming that the remaining measures do not have an II signal. Note that due to the exclusion of auto-correlation power spectra the statistical errors on cosmological parameters in this work are larger than those of other cosmic shear tomography analyses, even for our original data sets.

Excluding auto-correlations is of limited accuracy to control the II signal since we use a relatively dense binning, partially with large photometric errors, so that cross-correlations of adjacent photometric redshift bins would contain significant II terms as well. With realistic data one could in principle let the nulling be preceded by an II removal technique such as \citet{king02} who also take a purely geometric approach. However, the redshift-dependent weighting of galaxy pairs, on which the II removal is based, modifies the calculation of the projected cosmic shear measures such as (\ref{eq:limber}), which in turn entails a modification of the nulling weights. The improvements of the nulling technique we investigate in Sect.$\,$\ref{sec:controladjacent} will also constitute an efficient tool to control the II term.

\section{Improving the nulling performance}
\label{sec:optimization}

\subsection{Optimizing the nulling weights}
\label{sec:optnulling}

In the composition of the nulling weights (\ref{eq:defT}) one has the freedom to choose the specific redshift $\hat{z}_i$ within the initial bin at which the GI contribution is eliminated, as well as the referencing of redshifts $z_j$ to the background redshift bins. For convenience JS08 placed $\hat{z}_i$ at the center of the initial bin and identified $z_j$ with the lower boundary of bin $j$. Since this choice was fairly arbitrary, we seek to find a more appropriate referencing that leads to a minimum residual GI contamination.

\begin{figure*}[ht]
\begin{minipage}[c]{.75\textwidth}
\centering
\includegraphics[scale=.45]{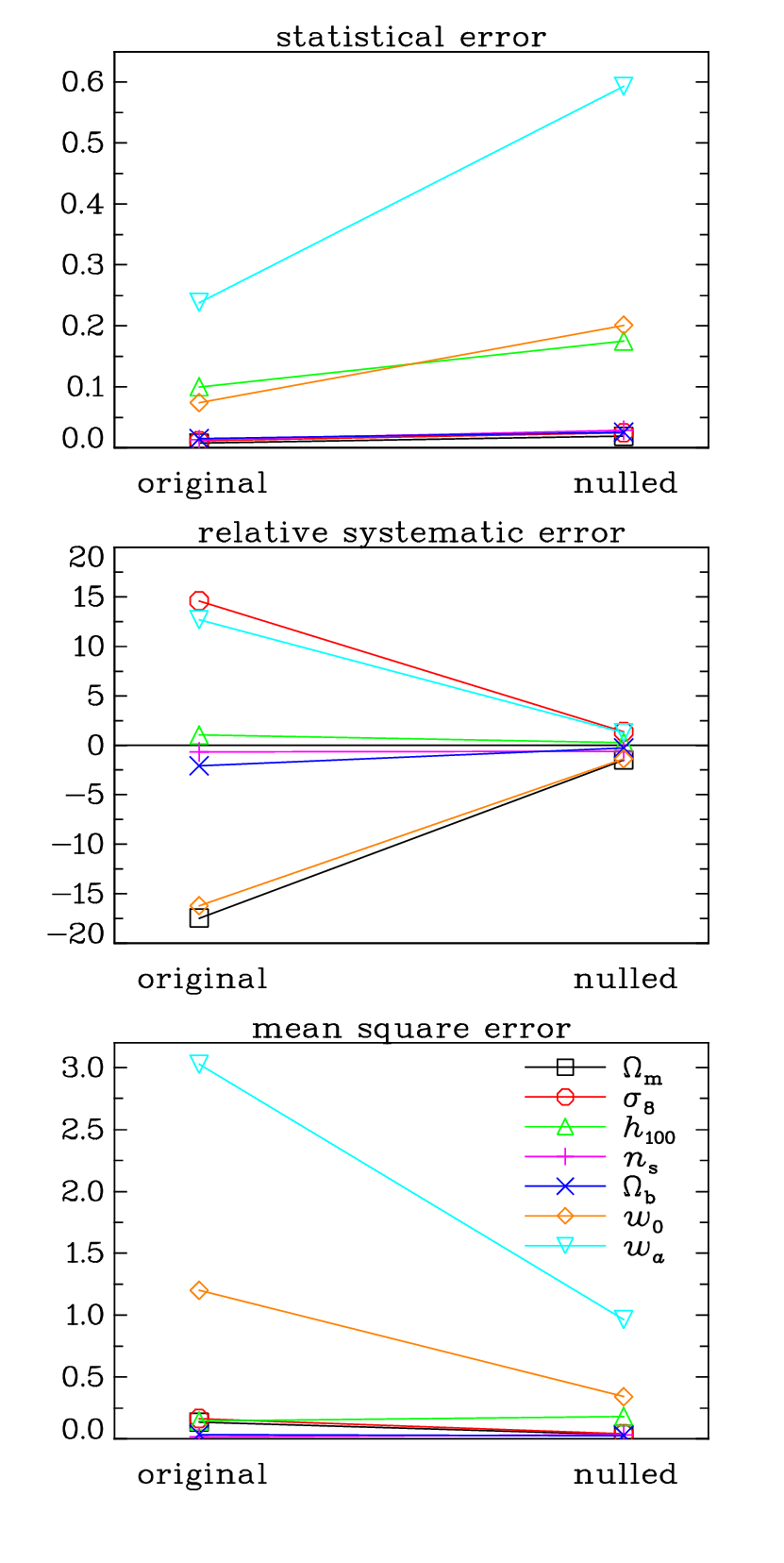}
\includegraphics[scale=.45]{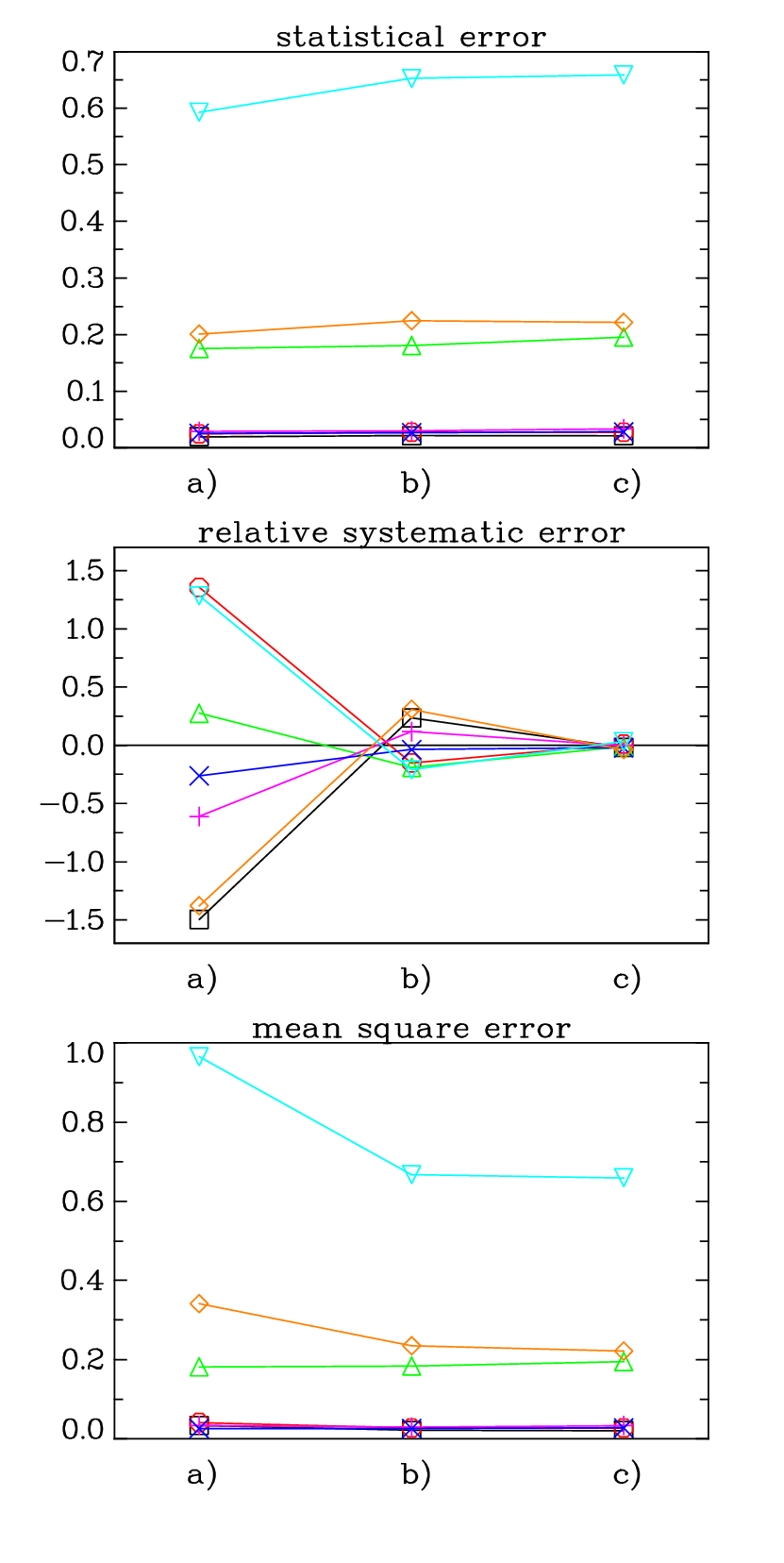}
\end{minipage}%
\begin{minipage}[c]{.25\textwidth}
\caption{Comparison of the performance of the different nulling weights. Shown are marginalized statistical errors $\sigma$ in the top panels, relative systematic errors $b_{\rm rel}$ in the center panels, and mean square errors $\sigma_{\rm tot}$ in the bottom panels. For the correspondence between considered parameters and line colors/symbols see the legend. \textit{Left column}: Change in errors from original to nulled data set, using the referencing to bin boundaries, i.e. variant (A). \textit{Right column}: Residual errors using the different nulling weights. (A) Referencing to bin boundaries; (B) Referencing to bin centers; (C) Nulling including detailed redshift information.}
\label{fig:comparenullingeffect}
\end{minipage}
\end{figure*}

A more natural choice is to position both the redshift of the initial bin $\hat{z}_i$ and the reference redshifts of the background bins at the center between the photometric redshift bin boundaries, denoted by $z^{(i)}_{\rm c}$. This setup does not require knowledge about the redshift probability distribution of each bin, although this information has to be available at high precision for future cosmic shear surveys. Hence, we furthermore define nulling weights that take redshift information into account. Re-examining (\ref{eq:PGI_approx}), one can drop the approximation of narrow redshift/distance probability distributions for the background bins, keeping the first equality of (\ref{eq:PGI_approx}). Thereby, instead of the comoving distance ratio $\br{1-\chi(\hat{z}_i)/\chi(z_j)}$, one directly uses the lensing efficiency, which is the average of this ratio, weighted by the redshift/distance probability distribution of the background photometric redshift bin. The zeroth-order nulling weight in (\ref{eq:defT}) is then given by ${{T'}_{[0]}^{(i)}}_j = g^{(j)}\br{\chi(\hat{z}_i)}$. For the remaining free redshift of the initial bin $\hat{z}_i$ we choose the median redshift of distribution $i$, a measure that contains information about the form of the distribution, but is robust against outliers.

\begin{table}[t]
\caption{Overview on nulling variants considered. The variants differ by the redshifts assigned to the foreground and background photometric redshift bins, and by the form of the zeroth-order weight function.}
\centering
\begin{tabular}[t]{cccc}
variant & foreground & background & $0^{\rm th}$ order weights\\
\hline
\hline
(A) & bin center & lower boundary & $1-\chi(\hat{z}_i)/\chi(z_j)$\\
(B) & bin center & bin center & $1-\chi(\hat{z}_i)/\chi(z_j)$\\
(C) & median redshift & bin center & $g^{(j)}\br{\chi(\hat{z}_i)}$\\
\end{tabular}
\label{tab:nullingvariants}
\end{table}

Hence, in total we are going to consider three different versions of nulling: (A) the \lq old\rq\ version of nulling with referencing to the lower boundaries of the background bins, a variant (B) where the background bins are identified with the bin centers $z^{(i)}_{\rm c}$ instead, and (C) the nulling that includes detailed redshift information via assigning the foreground bins to their median redshifts and using the comoving distance ratio, weighted by $p^{(j)}(\chi)$, as the zeroth-order nulling weight. The properties of these variants are summarized in Table \ref{tab:nullingvariants}.

In Fig.$\,$\ref{fig:comparenullingeffect} the performance of nulling with different nulling weights is shown. We plot the marginalized statistical error $\sigma(p_\mu) = \sqrt{(F^{-1})_{\mu\mu}}$ and the relative bias
\eq{
\label{eq:brel}
b_{\rm rel}(p_\mu) \equiv b(p_\mu)/\sigma_{\rm orig}(p_\mu)\;,
}
where $\sigma_{\rm orig}$ denotes the statistical error before nulling, for every cosmological parameter. Note that if we referred the bias after nulling to the statistical error after nulling, the usual loss of information due to nulling could cause a decrease in $b/\sigma$ even if the GI contamination remained completely unmodified. With the definition (\ref{eq:brel}), $b_{\rm rel}$ is an unambiguous measure of the relative importance of systematic errors in the data. Moreover, the mean square error (\ref{eq:sigmatot}) is given in the figure. Here and in the following, the seven parameters $\vek{p}=\bc{\Omega_{\rm m},\sigma_8,h_{100},n_{\rm s},\Omega_{\rm b},w_0,w_a}$ are considered in the Fisher matrix analysis. The data set is composed of power spectra for $N_z=10$ bins without photometric redshift errors, where the systematic stems from the linear alignment model, downscaled by a factor of five.

The left column of Fig.$\,$\ref{fig:comparenullingeffect} illustrates the change in errors due to nulling with the referencing used hitherto, i.e. variant (A). While the marginalized statistical errors increase by up to a factor of about three for the weakly constrained dark energy parameters, the bias drops from values of up to $17\,\sigma$ to numbers that are of the same order of magnitude as the original statistical errors, i.e. $b_{\rm rel} \approx 1$. For parameters that were strongly biased this leads to a considerable decrease in the mean square error, but $\sigma_{\rm tot}$ may also slightly increase if the systematic was subdominant already before nulling as is the case for the Hubble parameter.

In the right column of Fig.$\,$\ref{fig:comparenullingeffect} resulting errors for all three nulling variants are given. It is evident that the newly introduced versions (B) and (C) of nulling perform significantly better in removing the systematic. Variant (B) decreases the bias by at least a factor of three with respect to (A), reversing the sign of the bias for almost all parameters. This hints at using the reference redshifts of the nulling weights as free parameters to control the amount of bias allowed in the data, as will be further discussed in Sect.$\,$\ref{sec:conclusions}. Variant (C) nearly perfectly eliminates the GI contamination. Although the underlying data lacks photometric redshift errors, knowledge about the distributions $p^{(i)}(z)$ is still advantageous as e.g. the lowest and highest redshift bin are broad and largely asymmetric. Regarding statistical errors, the better a version is capable of removing the systematic, the less stringent parameter constraints become. However, the improved bias reduction clearly outweighs the marginal increase in statistical errors.

In summary, we propose to henceforth use nulling with referencing to the centers of photometric redshift bin divisions, i.e. variant (B), in absence of detailed information about redshift distributions, and else version (C) which exploits this knowledge. Both approaches will be considered in the following analyses.

\subsection{Cosmology-dependence of the nulling weights}
\label{sec:cosdep}

The nulling weights ${T_{[q]}^{(i)}}_j$ depend on those parameters of the cosmological model that enter the comoving distance in a non-trivial way, i.e. for our model assumptions $\Omega_{\rm m}$, $w_0$, and $w_a$. Since only ratios of comoving distances enter the nulling weights, there is no dependence on $h_{100}$ which enters the prefactor of (\ref{eq:wz}). If the relevant cosmological parameters chosen to compute the nulling weights are different from the true parameters of the data set, the performance of nulling may deteriorate. A grossly incorrect choice of nulling weights could in principle affect the lensing signal more than the GI term, which could then even cause a larger bias on parameters in the transformed data than in the original one.

\begin{figure}[t]
\centering
\includegraphics[scale=.5]{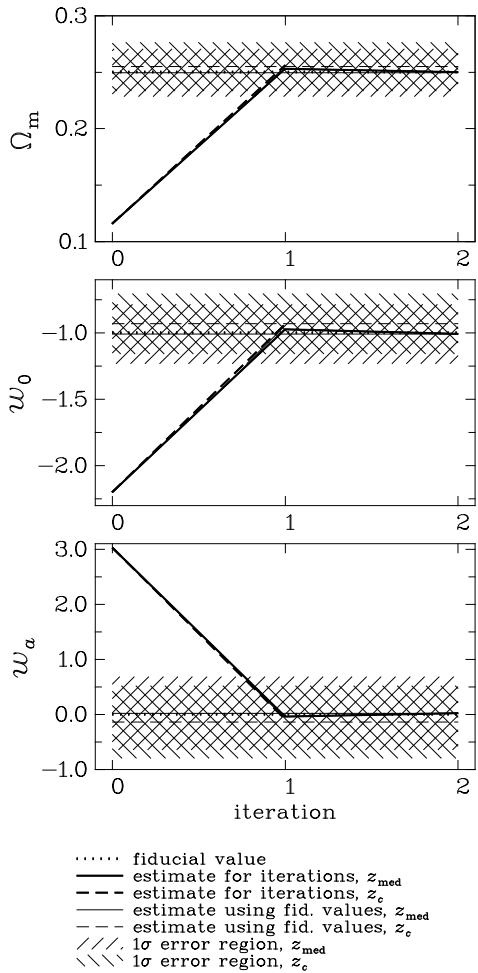}
\caption{Cosmology dependence of the nulling weights. The change in estimates for the cosmological parameters, entering the distance-redshift relation non-trivially, is plotted for different iteration steps. The estimates resulting from using variant (C) are shown as solid lines, those for variant (B) as dashed lines. Iteration 0 corresponds to the initial values for the parameters, in this case the results of the analysis of the unmodified data set. For reference, the estimates obtained by using the true underlying cosmology to compute the nulling weights are plotted as thin lines. The hatched regions around these lines signify the 1$\sigma$ error region. Note that variant (B) reaches an accuracy compatible to using the true cosmology already after one iteration while variant (C) takes two iterations.}
\label{fig:cosdep}
\end{figure}

Avoiding any a priori guesses of the true values of the relevant cosmological parameters, we explore the cosmology dependence of the nulling weights by taking the estimates from the analysis of the original data set as input cosmology for the computation of the ${T_{[q]}^{(i)}}_j$. As we use the linear alignment model (\ref{eq:GIlinalign}), the estimates $p_{\rm b}=p_{\rm f}+b$, where $p_{\rm f}$ is the true parameter value and $b$ is the bias, are far from the true values and beyond any decent a priori guess, so that this setup can be understood as a worst-case scenario. With the weights obtained this way, the nulled data can be analyzed, yielding another set of parameter estimates. This can then be taken as input for a refined set of nulling weights, thereby creating an iterative process which can be terminated when successive iterations yield stable parameter estimates.

In Fig.$\,$\ref{fig:cosdep} the results of this iteration process are shown for nulling variants (B) and (C), both showing a very similar behavior. The parameter estimates for iteration 0 correspond to the estimates of the analysis of the original data set. Given these largely incorrect input parameters, nulling is still able to reduce the bias due to intrinsic alignment to a level close to the one when using the true cosmology as input. Already after the first iteration step the residual bias is considerably smaller than the statistical errors. After at most two iterations, the results for the residual bias are indistinguishable from those with the correct input parameters.

Hence, the dependence of the nulling weights on cosmology is only weak, being solely due to geometrical terms. Consequently, nulling is robust against an incorrect initial guess for cosmological parameters needed to compute the nulling weights. For a consistency check, the iterative procedure outlined above can be performed on the data. In the remainder of this work we will use the true cosmology to calculate the nulling weights for reasons of simplicity.

\section{Influence of redshift information on nulling}
\label{sec:zinfo}

\subsection{Redshift binning}

First, we investigate the performance of nulling as a function of the number of photometric redshift bins the survey is divided into. The larger $N_z$, the better (\ref{eq:zwangsbed_discrete}) is an approximation of (\ref{eq:zwangsbed}), so that the GI removal is expected to work more efficiently. Furthermore, since nulling eliminates the contribution to the lensing signal of the background objects only at a single redshift, more concentrated redshift probability distributions are nulled more accurately, given an appropriately chosen redshift $\hat{z}_i$ within the initial bin. At the same time, less statistical information is lost because the entries of the transformed data vector, which are removed in the process of nulling, contain less independent information if the redshift distributions have a smaller spacing.

In search for a single quantity that measures an overall power of a data set to constrain cosmological parameters we define the average statistical power as
\eq{
\label{eq:averageF}
\bar{F} \equiv \bc{ \det \br{F_{\mu\nu}} }^{\frac{1}{2\,N_p}}\;,
}
where $N_p$ is the number of parameters considered, i.e. the dimension of the Fisher matrix. This measure is motivated by the fact that the determinant of the Fisher matrix is inversely proportional to the volume of the $N_p$-dimensional error ellipsoid in parameter space. If errors are not correlated, $\bar{F}^2$ reduces to the geometric mean of the inverse square errors. In addition, we introduce an average relative bias
\eq{
\label{eq:averagebias}
\bar{b} \equiv \sqrt{ \frac{1}{N_p}\; \sum_{\mu=1}^{N_p}\; \frac{b^2(p_\mu)}{\sigma_{\rm orig}^2(p_\mu)} } = \sqrt{ \frac{1}{N_p}\; \sum_{\mu=1}^{N_p}\; b_{\rm rel}^2(p_\mu) }\;,
}
which is the root mean square of the ratio of the systematic over the statistical error before nulling over all considered parameters. We refer to the performance of nulling via the ratios
\eq{
\label{eq:defratios}
r_F \equiv \frac{\bar{F}_{\rm null}}{\bar{F}_{\rm orig}}\;; ~~~~r_b \equiv \frac{\bar{b}_{\rm null}}{\bar{b}_{\rm orig}}
}
of $\bar{F}$ and $\bar{b}$ after (\lq null\rq) and before (\lq orig\rq) nulling, respectively. For a good performance of nulling, $r_F$ should tend to one, i.e. the nulled data constrains parameters as well as the original one, whereas $r_b$ tends to zero, which corresponds to a complete elimination of the systematic.

\begin{figure}[t]
\centering
\includegraphics[scale=.61,angle=270]{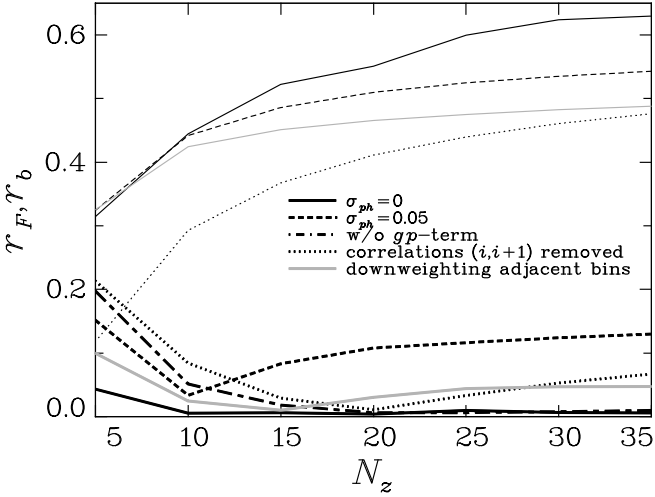}
\caption{Ratios $r_F$ and $r_b$ as a function of the number of photometric redshift bins $N_z$. Thin curves represent $r_F$, thick curves $r_b$. Results for zero photometric redshift error are given as solid black lines; results for $\sigma_{\rm ph}=0.05$ are plotted as dashed lines. For the case $\sigma_{\rm ph}=0.05$, $r_b$ is also plotted without the $gp$-term included in the calculation of the systematic, see the dot-dashed line. Since only the systematic signal is manipulated, the statistical signal in this case is still given by the dashed line. Dotted lines represent $r_F$ and $r_b$ if correlations of adjacent bins, i.e. bin combinations $(ij)$ with $j=i+1$, are excluded. Incorporating the downweighting scheme for correlations of adjacent bins introduced in Sect.$\,$\ref{sec:controladjacent} produces the gray solid curves. The two latter sets of curves were also obtained for $\sigma_{\rm ph}=0.05$. Note that the black solid and the dot-dashed lines are very close to zero for $N_z>10$ and $N_z>20$, respectively.}
\label{fig:nzbias}
\end{figure}

Figure \ref{fig:nzbias} shows results for the ratios $r_F$ and $r_b$ for different $N_z$, both without photometric redshift errors and for $\sigma_{\rm ph}=0.05$. In this section the linear alignment model is used as the systematic, downscaled by a factor of five. For five redshift bins $\bar{F}_{\rm null}$ is only about a third of $\bar{F}_{\rm orig}$, but $r_F$ rises, first strongly and then with an increasingly shallow slope for larger $N_z$. This development is mostly based on the improving performance of nulling since for a cosmic shear tomography data set statistical errors only marginally decrease for $N_z \geq 5$ (see e.g. \citealp{hu99,simon04,ma05,bridle07}; JS08).

Introducing a photometric redshift dispersion of $\sigma_{\rm ph}=0.05$, one finds that, for small $N_z$, $r_F$ increases in the same way as in the case without photometric redshift errors. As soon as the size of the redshift bins attains the same order as the width of the dispersion $\sigma_{\rm ph}(1+z)$, less additional redshift information becomes available to constrain parameters. Since nulling, like other techniques that deal with the control of intrinsic alignments \citep[e.g.][]{bridle07}, requires more precise redshift information, the curve for $r_F$ levels off.

Even for only five bins in redshift, nulling is capable of reducing the average bias $\bar{b}$ by more than $95\,\%$ for perfect redshift information. For $N_z \geq 10$, less than $1\,\%$ of the average bias remains. If a more realistic photometric redshift dispersion is present in the data, $r_b$ significantly degrades to approximately 0.15 for $N_z=5$. For ten photometric redshift bins a minimum value of $r_b \approx 3.5\,\%$ is achieved before this ratio increases again for more bins, meaning that the treatment of the systematic worsens in spite of the improvement of redshift information due to the finer division of photometric redshifts. This apparent contradiction requires a more thorough investigation and will be addressed in Sect.$\,$\ref{sec:controladjacent}.

\subsection{Minimum information loss}

\begin{figure}[t]
\centering
\includegraphics[scale=.58,angle=270]{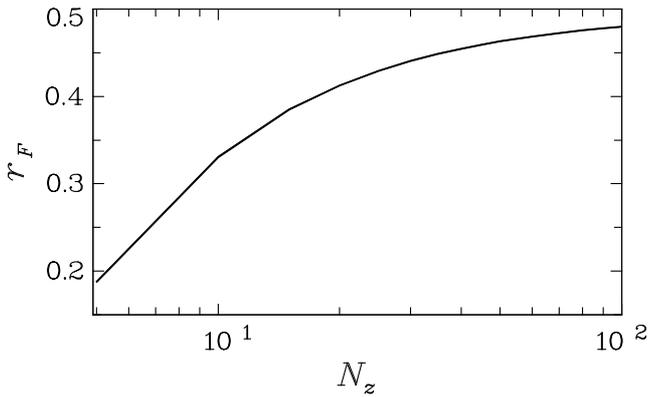}
\caption{Ratio $r_F$ as a function of the number of photometric redshift bins $N_z$. This result has been obtained by means of a simplified Fisher matrix calculation, placing galaxies at fixed redshifts and neglecting cosmic variance in the covariance. For large $N_z$ the increase in $r_F$ is slower than logarithmic.}
\label{fig:inevloss}
\end{figure}

Given ideal spectroscopic redshift information, equivalent to considering the limit $N_z \rightarrow \infty$, it would be possible to precisely eliminate the GI contamination at a given redshift, see (\ref{eq:PGI_approx}), so that $r_b$ tends to zero in absence of photometric redshift errors, as is indeed the case. However, the curves for $r_F$ in Fig.$\,$\ref{fig:nzbias} apparently indicate that the full statistical information is not regained in this limit, i.e. $r_F$ does not tend to unity. We investigate this further by calculating $r_F$ out to larger $N_z$, assuming a simplified model with infinitesimally narrow redshift bins,
\eq{
\label{eq:sheetbin}
p^{(i)}(z)=\delta_{\rm D}(z-z_i)\;,
}
and a covariance that contains only shot noise. The resulting curve, shown in Fig.$\,$\ref{fig:inevloss}, increases slower than logarithmically as a function of $N_z$, so that one can expect that indeed nulling inevitably reduces the statistical power of a data set, even when spectroscopic redshifts would be available.

To illustrate this effect, consider again the continuous, integral version of (\ref{eq:defT}), still in the limit of perfect redshift information. Choosing the zeroth-order nulling weight proportional to $1-\chi_i/\chi_j$, see (\ref{eq:defT}), one can write the corresponding transformed power spectrum as
\eqa{
\label{eq:zeroordernulledps1}
\Pi_{[0]}(\ell,\chi_i) &\propto& \int_{\chi_i}^{\chi_{\rm hor}} \dd \chi_j\; \br{1 - \frac{\chi_i}{\chi_j}}\; P_{\rm GG}(\ell,\chi_i,\chi_j)\\ \nn
&\propto& \int_{\chi_i}^{\chi_{\rm hor}} \dd \chi_j\; \br{1 - \frac{\chi_i}{\chi_j}}\; \int_0^{\chi_i} \dd \chi\; \br{1 - \frac{\chi}{\chi_i}}\; \br{1 - \frac{\chi}{\chi_j}}\\ \nn
&& \hspace*{3cm} \times \bc{1+z(\chi)}^2 P_{\delta\delta} \br{\frac{\ell}{\chi},\chi}\;,
}
where in order to arrive at the second equality, the lensing power spectrum for spectroscopic redshifts has been obtained by inserting (\ref{eq:sheetbin}) into (\ref{eq:limber}). Note that the upper limit in the integration over $\chi$ changes from $\chi_{\rm hor}$ to $\chi_i$ because the lensing efficiency, here written as $1-\chi/\chi_i$, vanishes for $\chi > \chi_i$. Rearranging the terms, one arrives at
\eqa{
\label{eq:zeroordernulledps}
\Pi_{[0]}(\ell,\chi_i) &\propto& \int^{\chi_i}_0 \dd \chi\; \br{1 - \frac{\chi}{\chi_i}}\; \bar{g}(\chi) \bc{1+z(\chi)}^2 P_{\delta\delta} \br{\frac{\ell}{\chi},\chi}\\ \nn
\mbox{with} ~~~  \bar{g}(\chi) &\equiv& \int_{\chi_i}^{\chi_{\rm hor}} \dd \chi_j\; \br{1 - \frac{\chi_i}{\chi_j}}\; \br{1 - \frac{\chi}{\chi_j}}\;.
}
Comparing (\ref{eq:zeroordernulledps}) to (\ref{eq:limber}), one finds that the term $\bar{g}(\chi)$ is formally equivalent to the lensing efficiency of the background distribution\footnote{For perfect correspondence the lower limit of the integral over $\chi_j$ should be $\chi$ instead of $\chi_i$. However, the nulling weight given as $1-\chi_i/\chi_j$ has to vanish for $\chi_j<\chi_i$, and at the same time the outer integral ensures $\chi<\chi_i$.}, the term $1-\chi_i/\chi_j$ acting analogously to a distance probability distribution of galaxies. Thus, this \lq background distribution\rq\ of the transformed power spectrum is broad, extending from the position of the foreground bin at $\chi_i$ to the maximum distance $\chi_{\rm hor}$. Since the zeroth-order nulled power spectra are removed from the data set, it is this integrated redshift information for all foreground bin positions $\chi_i$ that is necessarily lost due to nulling.

\subsection{Intrinsic alignment contamination from adjacent bins}
\label{sec:controladjacent}

The increase in $r_b$ for large $N_z$ in the case $\sigma_{\rm ph}=0.05$, as seen in Fig.$\,$\ref{fig:nzbias}, can be explained by inspecting (\ref{eq:limberGI}). To produce a GI effect, the intrinsic alignment has to act on the foreground galaxy while the background galaxy is lensed. Hence, the GI signal should stem from the first term in (\ref{eq:limberGI}), whereas the second term that contains $g^{(i)}(\chi)~p^{(j)}(\chi)$ with $i<j$ vanishes if the redshift probability distributions are disjoint, see (\ref{eq:PGI_approx}). We refer to the latter expression as the $gp$-term hereafter. This term can yield a contribution to the systematic in case the distributions overlap such that the true position of a galaxy from the background population is in front of galaxies from the foreground distribution. The contribution to the GI signal by swapped galaxy positions is not accounted for by nulling and produces a residual systematic.

To quantify the effect caused by the $gp$-term, we compute the average bias for the same model of the three-dimensional GI power spectrum, but now with the $gp$-term removed from (\ref{eq:limberGI}). The resulting ratio $r_b$ is plotted in Fig.$\,$\ref{fig:nzbias} as well. While this curve shows a similar behavior than the one for the systematic with $gp$-term for $N_z \leq 10$, it does not follow the turnaround and continues to decrease for larger $N_z$ down to values of $r_b$ obtained for data without photometric redshift errors, as expected. Thus, the increase in $r_b$ of the data with $\sigma_{\rm ph}=0.05$ for $N_z > 10$ can indeed be explained by the contamination due to the $gp$-term.

The $gp$-term cannot be quantified in detail as it depends explicitly on the form of the matter-intrinsic shear power spectrum, see (\ref{eq:limberGI}). However, it is produced by an overlap of the redshift distributions of foreground and background distributions, so that the $gp$-term can be controlled by removing or downweighting bin combinations with a large overlap in redshift, in particular adjacent photometric redshift bins. For instance, one can simply exclude power spectra for bins $(ij)$ with $j=i+1$ from the analysis, which results in the dotted curves given in Fig.$\,$\ref{fig:nzbias}. Indeed the contamination by the $gp$-term is suppressed, producing merely a less significant increase in $r_b$ for $N_z>20$, but the statistical power decreases dramatically due to the removal of all power spectra with $j=i+1$.

To alleviate this effect, we propose to downweight adjacent redshift bin combinations. According to (\ref{eq:require_higherorder}), increasing an entry in the zeroth-order nulling weight implies a lower value in the corresponding entries of the higher-order weights. Hence, a manipulation of the zeroth-order weights can be used to downweight certain power spectra in the process of nulling. We introduce the following modified weights
\eqa{
\label{eq:gaussianweights}
{{T'}^{{\rm w},\,(i)}_{[0]}}_j &\equiv& w_{ij}\; {{T'}^{(i)}_{[0]}}_j  ~~~\mbox{with}\\ \nn
w_{ij} &=& 1 + \exp \bc{- \br{ \frac{\hat{z}_j - \hat{z}_i}{\sigma_{\rm ph} \br{1+\hat{z}_i}} }^2}\;.
}
To motivate this choice, consider that for $j \gg i$ one gets $w_{ij} \approx 1$, so that in the regime where the $gp$-term is unimportant the original weights are reproduced. Moreover, $w_{ii}=2$, which is in agreement with the fact that the $gp$-term is equal to the first term in (\ref{eq:limberGI}) for auto-correlations (note however that auto-correlations are excluded from the analysis anyway). The width of the Gaussian in (\ref{eq:gaussianweights}) is in principle arbitrary, but here conveniently chosen to scale with the width of the photometric redshift bins. 

Therefore, the $w_{ij}$ are expected to follow the redshift dependence of the $gp$-term, so that the higher-order nulling weights $\vek{T}^{{\rm w},\,(i)}_{[q]}$ with $q \geq 1$ efficiently downweight its contribution. Note that the modification of the nulling weights is done before normalization such that the vectors $\vek{T}^{{\rm w},\,(i)}_{[q]}$ still have unit length. As an aside, the weighting scheme (\ref{eq:gaussianweights}) would also contribute to the downweighting of contaminations by the II term.

Applying this Gaussian weighting scheme to the nulling procedure, one obtains the gray curves of Fig.$\,$\ref{fig:nzbias}. While for a small number of redshift bins $r_F$ is similar to the case where all power spectra except auto-correlations were used, the curve approaches the results for the case with power spectra of adjacent bins removed for large $N_z$. This means that for small $N_z$ the overlap between redshift bins is marginal, so that the weighting has only little effect, whereas for many bins power spectra with $j=i+1$ are largely downweighted such that removing them produces similar results. The Gaussian weighting ensures that $r_b \lesssim 5\,\%$ for all $N_z>10$. We will further consider the performance of this weighting scheme in Sect.$\,$\ref{sec:outliers}.

The best binning in photometric redshifts in terms of nulling performance does not only depend on the number of bins $N_z$, but to a certain extent also on the choice of bin boundaries. The optimal positions of bin boundaries are determined by the detailed form of the relation between photometric and true, spectroscopic redshifts, which is specific to each survey and thus shall not be further assessed here.

\section{Influence of photometric redshift uncertainty}
\label{sec:photozerrors}

\subsection{Photometric redshift errors}

This section deals with the dependence of nulling on the photometric redshift dispersion $\sigma_{\rm ph}$, in absence of catastrophic outliers. The number of photometric redshift bins is kept at $N_z=10$ for the remainder of this work, mainly for computational reasons. Future cosmic shear surveys, relying on precise redshift information and a large number of galaxy detections, will allow for considerably more photometric redshift bins, which may be advantageous in terms of nulling, see the foregoing section.

\begin{figure}[t]
\centering
\includegraphics[scale=.61]{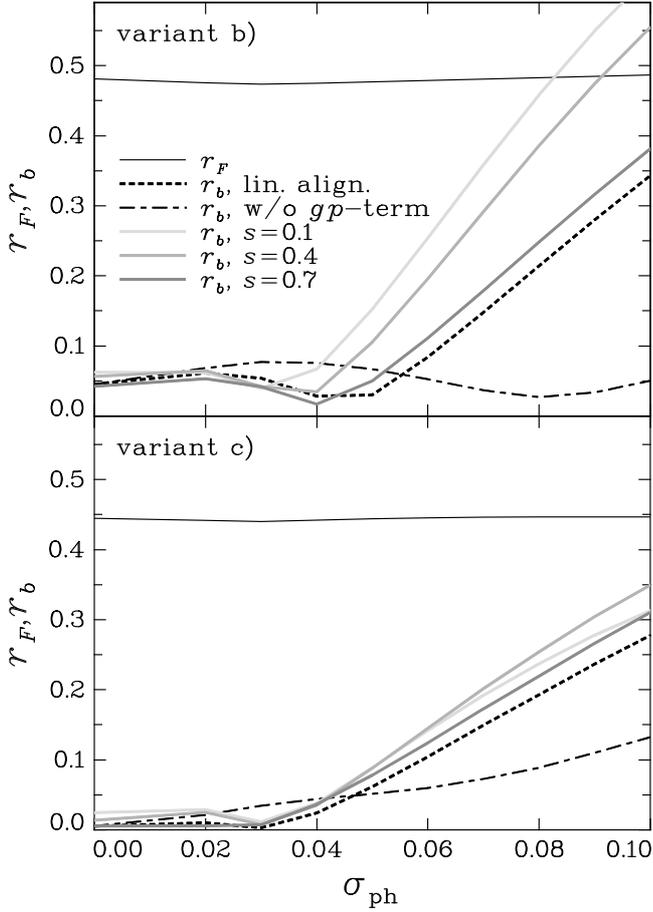}
\caption{\textit{Top panel}: Ratios $r_F$ and $r_b$ as a function of photometric redshift dispersion $\sigma_{\rm ph}$. The nulling has been performed by using variant (B), and the linear alignment model, downscaled by a factor of five, has been employed as systematic. Solid black curves correspond to $r_F$ while $r_b$ for the linear alignment model as systematic is given as black dashed curve. The values of $r_b$ for the same model, but with the $gp$-term removed from the GI power spectrum calculation, is given as dot-dashed line. The gray curves show $r_b$ for the GI power-law models, where the different gray-scales stand for different slopes $s$ as given in the legend. \textit{Bottom panel}: Same as above, but using nulling variant (C).}
\label{fig:sigmaphot_total}
\end{figure}

\begin{figure}[t]
\centering
\includegraphics[scale=.58]{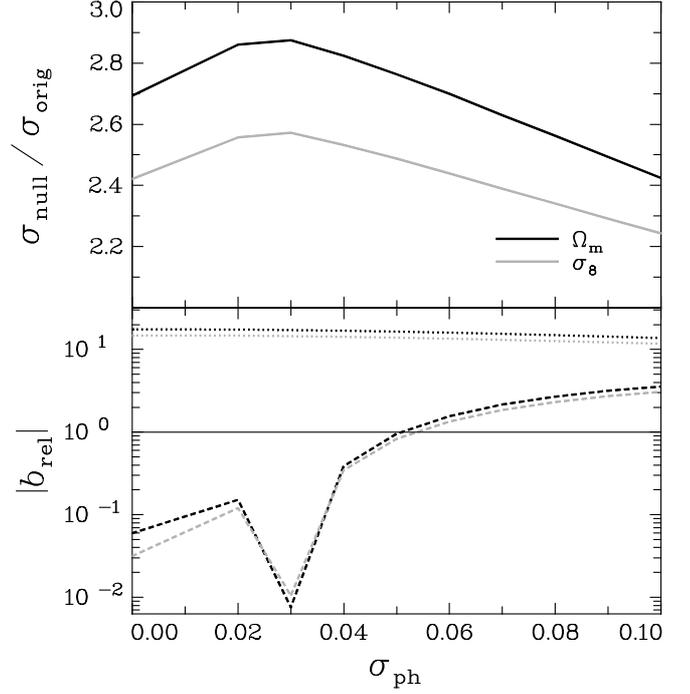}
\caption{Performance of nulling as a function of photometric redshift dispersion $\sigma_{\rm ph}$. The nulling has been done using variant (C), and the linear alignment model, downscaled by a factor of five, has been employed as systematic. Shown are the results for the parameters $\Omega_{\rm m}$ as black curves, and for $\sigma_8$ as gray curves. \textit{Top panel}: Ratio of the marginalized statistical errors after and before nulling. \textit{Bottom panel}: Relative bias $b_{\rm rel}$. Dotted curves correspond to $b_{\rm rel}$ before nulling; dashed curves to $b_{\rm rel}$ after nulling. The solid line marks values of $b_{\rm rel}$ for which the marginalized statistical errors equal the bias. Note the logarithmic scaling of the ordinate axis.}
\label{fig:sigmaphot_stat}
\end{figure}

In Fig.$\,$\ref{fig:sigmaphot_total} $r_F$ is plotted as a function of $\sigma_{\rm ph}$ while in Fig.$\,$\ref{fig:sigmaphot_stat}, upper panel, the ratios of the marginalized statistical errors before and after nulling are given for the parameters $\Omega_{\rm m}$ and $\sigma_8$ individually. The curves for the other cosmological parameters vary considerably in magnitude, but otherwise show the same characteristics as the ones depicted. The ratio $r_F$ decreases only very weakly with increasing $\sigma_{\rm ph}$ for both nulling variants (B) and (C), taking values between 0.44 and 0.48, because splitting the range of redshifts between 0 and 3 into 10 photometric redshift bins does not lead to a significant degrading of redshift information, even for $\sigma_{\rm ph}=0.1$. In contrast to this, the ratio of the marginalized errors of individual cosmological parameters does vary with $\sigma_{\rm ph}$, but changes are smaller than about $10\,\%$. The statistical errors of both the original and the nulled data set increase for larger photometric redshift errors similarly, but the error of the nulled set starts to do so already at smaller $\sigma_{\rm ph}$, thereby producing a peak at $\sigma_{\rm ph}\approx0.03$ in both curves in Fig.$\,$\ref{fig:sigmaphot_stat}. Marginalized errors for each of the seven considered parameters are a factor of roughly two to three larger for the nulled data.

As is evident from Fig.$\,$\ref{fig:sigmaphot_total}, lower panel, nulling using variant (C) is capable of reducing the average bias caused by the linear alignment model by more than a factor of 50 for $\sigma_{\rm ph} \lesssim 0.04$. Looking at the effect on the bias of individual parameters in Fig.$\,$\ref{fig:sigmaphot_stat}, lower panel, one sees that the systematic is suppressed by more than 2 orders of magnitude for small $\sigma_{\rm ph}$. In spite of the strong intrinsic alignment signal, the bias is kept subdominant up to $\sigma_{\rm ph} \approx 0.05$. The drop in $r_b$ at $\sigma_{\rm ph} \sim 0.03$ is also visible in Fig.$\,$\ref{fig:sigmaphot_total} and can be traced back to a sign change in the residual bias for several parameters, among them $\Omega_{\rm m}$ and $\sigma_8$.

For larger redshift dispersions, $r_b$ shows an approximately linear increase, which can only partially be ascribed to the contamination by the $gp$-term as can be concluded from comparing with the curve for the linear alignment model without $gp$-term. The rise in $r_b$ is caused by two effects that are visible in Fig.$\,$\ref{fig:sigmaphot_stat}. First, the strong relative bias in $\Omega_{\rm m}$ and $\sigma_8$ for the original data set starts to slowly decrease for $\sigma_{\rm ph} \gtrsim 0.02$, predominantly because the statistical errors rise due to the degrading information content in the line-of-sight direction. Second, the residual bias after nulling increases as a function of $\sigma_{\rm ph}$ and starts to attain values of the same order as the statistical errors, i.e. $|b_{\rm rel}|\sim1$, at just about $\sigma_{\rm ph} \approx 0.05$. The part of this degradation that cannot be traced back to the effect by the $gp$-term has to stem from the incorrect assessment of the redshift dependence of the GI signal, either due to the approximations inherent to the derivations of nulling or the suboptimal placement of the redshift at which the signal is nulled.

Figure \ref{fig:sigmaphot_total} also shows $r_b$ for the power-law GI model with varying slopes. The behavior of $r_b$ as a function of $\sigma_{\rm ph}$ is in very good agreement with the results for the linear alignment model, $r_b$ reaching about 0.03 for $\sigma_{\rm ph} \lesssim 0.04$, and up to $30\,\%$ higher values for $\sigma_{\rm ph} = 0.1$ in comparison with the linear alignment model. This suggests that at least the orders of magnitude of our results as well as the general conclusions drawn from a particular GI model used in this work can be taken to robustly estimate the effects of a realistic GI contamination.

Moreover, Fig.$\,$\ref{fig:sigmaphot_total}, upper panel, illustrates the performance of nulling using variant (B), i.e. renouncing on information about the form of the redshift probability distributions, and placing the redshift at which the signal is nulled at the centers of the photometric redshift bins $z^{(i)}_{\rm c}$, respectively. This version of nulling is capable of retaining marginally more information in the data, in particular for small $\sigma_{\rm ph}$. For high quality redshift information the reduction in bias is worse, $r_b$ doubling approximately compared to variant (C). Again at $\sigma_{\rm ph} \sim 0.04$, $r_b$ starts to increase, but more steeply, so that for $\sigma_{\rm ph} > 0.04$ nulling quickly becomes rather inefficient. As for variant (C), the curves for $r_b$ of the different GI models agree well in their functional form, but yield largely different amplitudes. It is striking that the curve calculated without the $gp$-term does not feature a distinct increase for large $\sigma_{\rm ph}$. This suggests that variant (B), when combined with the weighting scheme of Sect.$\,$\ref{sec:controladjacent}, could perform well also for larger photometric redshift errors, as we will investigate in Sect.$\,$\ref{sec:outliers}.

\subsection{Analyzing optimal nulling redshifts}

The construction of nulling weights allows for a certain freedom in the choice of redshifts, which the photometric redshift bins are assigned to. We wish to investigate which choice of redshifts $\hat{z}_i$, i.e. those redshifts where the signal is nulled, is optimal in the sense that the resulting zeroth-order nulling weights (\ref{eq:defT}) best reproduce the redshift dependence of the GI signal, and thus effectively remove the systematic. The procedure to find such optimal nulling redshifts, denoted by $z_{\rm null}$, is outlined in the following. We emphasize that the calculation of $z_{\rm null}$ merely constitutes a diagnostic tool, inapplicable to data, since the GI systematic has to be known exactly to do this.

\begin{figure}[t]
\centering
\includegraphics[scale=.6,angle=270]{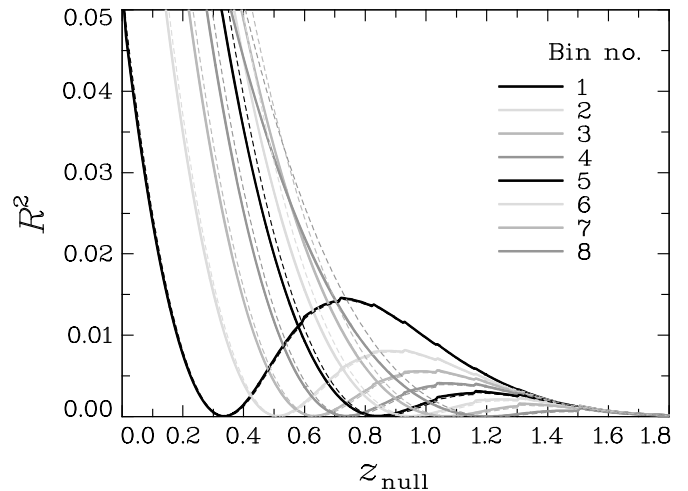}
\caption{Least squares sum $R^2$ as a function of nulling redshift $z_{\rm null}$. The results for photometric redshift bins one to eight correspond to the suite of gray-scale curves as given in the legend. Thin dashed lines represent the results for $R^2$ obtained when calculating the power spectrum without $gp$-term. Since we used $\sigma_{\rm ph}=0.05$ to produce this data, the minima of the latter curves are slightly offset. The local minima of these curves correspond to the optimal nulling redshifts $z_{\rm null}$ plotted in Fig.$\,$\ref{fig:znull_phot}. Note that $R^2$ at the local minima is close to, but always larger than zero.}
\label{fig:znull_leastsquares}
\end{figure}

Judging from (\ref{eq:PGI_approx}) and the considerations in Sect.$\,$\ref{sec:optnulling}, using the lensing efficiency $g^{(j)}\br{\chi(\hat{z}_i)}$ as zeroth-order nulling weight is most effective in case of precise redshift information. In fact, in the limit of spectroscopic redshifts $g^{(j)}\br{\chi(\hat{z}_i)}$ matches the redshift dependence of the GI signal perfectly. In the approximation of infinitesimally narrow redshift probability distributions for the photometric redshift bins with lower median redshift, i.e. the initial bins, the redshifts $\hat{z}_i$ would mark the position, at which the GI signal would be perfectly removed. In reality, the photometric redshift bins $i$ have finite size as do the corresponding distributions of true redshifts $p^{(i)}(z)$. The nulling redshift $\hat{z}_i$ is not fully specified anymore and has to be chosen appropriately. One reasonable choice is the median redshift of bin $i$, which corresponds to nulling variant (C). In this section we treat the $\hat{z}_i$ as free parameters and determine an optimal value $z_{\rm null}$.

\begin{figure}[t]
\centering
\includegraphics[scale=.6]{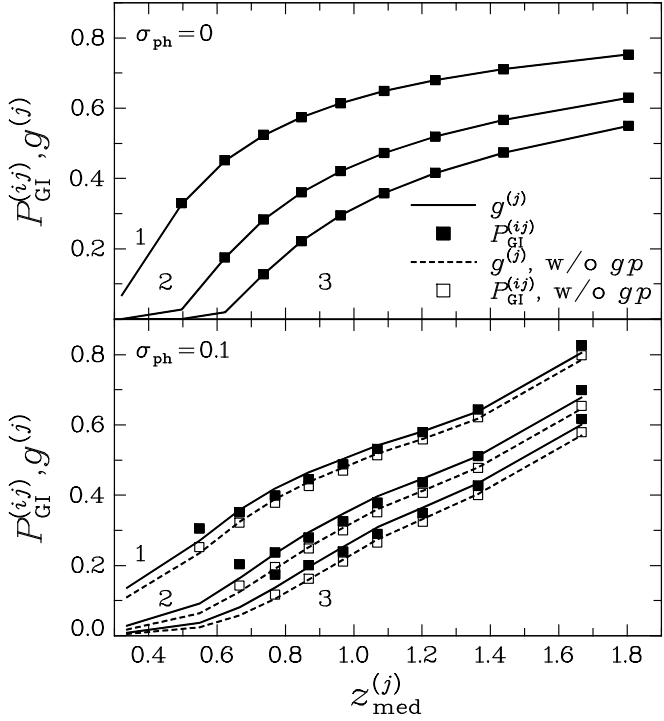}
\caption{Determination of the optimal nulling redshift. \textit{Top panel}: Results for $\sigma_{\rm ph}=0$. The filled squares display the redshift dependence of the GI power spectrum, i.e. $A_P P_{\rm GI}^{(ij)}(\ell)$ are plotted for different background bins $j$ and fixed $i$ and $\ell$. The lines correspond to the lensing efficiencies $g^{(j)}\br{\chi(\hat{z}_i)}$ for the best-fitting $\hat{z}_i$, respectively. The values for bin $j$ of both lensing efficiencies and power spectra have been assigned to the median redshift of this bin, linearly interpolating in between for $g^{(j)}\br{\chi(\hat{z}_i)}$. The numbers alongside the curves mark the initial bin number $i$. \textit{Bottom panel}: Same as above, but for $\sigma_{\rm ph}=0.1$. Here we plot in addition the results obtained by excluding the $gp$-term from the calculation of the GI signal as dashed curves and open squares, respectively.}
\label{fig:plotpgi}
\end{figure}

Hence, we aim at determining $\hat{z}_i$ such that $g^{(j)}\br{\chi(\hat{z}_i)}$ fits $P_{\rm GI}^{(ij)}(\ell)$ best since then nulling completely removes the intrinsic alignment signal with $g^{(j)}\br{\chi(\hat{z}_i)}$ as zeroth-order weight. To this end, we compute the best fitting lensing efficiency, using the least squares sum of all background bins $j$,
\eq{
\label{eq:leastsquares}
R^2\br{A_P,\hat{z}_i} = \sum_{j=i+1}^{N_z} \br{A_P P_{\rm GI}^{(ij)}(\ell) - g^{(j)}\br{\chi(\hat{z}_i)}}^2\;,
}
where the initial bin $i$ and the angular frequency $\ell$ are fixed. As default, we employ the values of $P_{\rm GI}^{(ij)}(\ell)$ for the central angular frequency bin, i.e. the bin with index $N_\ell/2$, which corresponds to $\ell \approx 414$. We warn that this is a crude approximation as the three-dimensional intrinsic alignment power spectrum varies significantly over the range of the integral in (\ref{eq:limberGI}). The redshift-independent part of the dependence of the GI power spectrum on $\ell$ can be absorbed into the free scaling $A_P$. The remaining $\ell$-dependence is accounted for by determining $z_{\rm null}$ for different angular frequencies, see Fig.$\,$\ref{fig:znull_phot} below. 

Since differences in the amplitude of $P_{\rm GI}^{(ij)}(\ell)$ and $g^{(j)}\br{\chi(\hat{z}_i)}$ are not of interest, the dependence of $R^2$ on the scaling is eliminated by calculating the extremal $A_P$ from the condition $\partial R^2/\partial A_P = 0$, yielding
\eq{
\label{eq:extremalscaling}
A_P = \frac{\sum_{j=i+1}^{N_z} g^{(j)}\br{\chi(\hat{z}_i)} P_{\rm GI}^{(ij)}(\ell)}{\sum_{j=i+1}^{N_z} \br{P_{\rm GI}^{(ij)}(\ell)}^2}\;.
}
Now $R^2$ is computed for a wide range of $\hat{z}_i$, making use of the fact that (\ref{eq:extremalscaling}) reduces the problem to a one-dimensional minimization. The value of $\hat{z}_i$ that corresponds to the minimum least squares is then set as the optimal nulling redshift $z_{\rm null}$.

In Fig.$\,$\ref{fig:znull_leastsquares} the least squares sum $R^2$ is plotted as a function of the $\hat{z}_i$ for a data set with $\sigma_{\rm ph}=0.05$, using the downscaled linear alignment model to compute the GI power spectrum. Note that for high redshifts $\hat{z}_i$, the lensing efficiency tends to zero, thereby implying an extremal value of $A_P=0$. Thus, the least squares go to zero for high redshifts because a GI power spectrum, scaled to zero, fits a vanishing lensing efficiency perfectly. The optimal nulling redshift is therefore extracted from the well-defined local minima of $R^2$, which can be clearly seen in Fig.$\,$\ref{fig:znull_leastsquares}.

The procedure to compute $z_{\rm null}$ is illustrated by Fig.$\,$\ref{fig:plotpgi}. The redshift dependence of the GI power spectra for initial bins 1 to 3, and the corresponding best-fit lensing efficiencies are plotted, referring the values for bin $j$ of both quantities to the median redshift of distribution $p^{(j)}(z)$.\footnote{This referring is merely for illustrative purposes and not part of the procedure outlined above.} The curves corresponding to the lensing efficiency are obtained via linear interpolation of the set of $g^{(j)}\br{\chi(\hat{z}_i)}$ with $j=i+1,\,..\,,N_z$. For the case without photometric redshift errors, nulling redshifts can be found such that the resulting lensing efficiencies almost exactly fit the redshift dependence of the GI power spectrum, so that in this case the approximation of infinitesimally narrow initial bins has little negative influence on the nulling performance.

\begin{figure}[t]
\centering
\includegraphics[scale=.6]{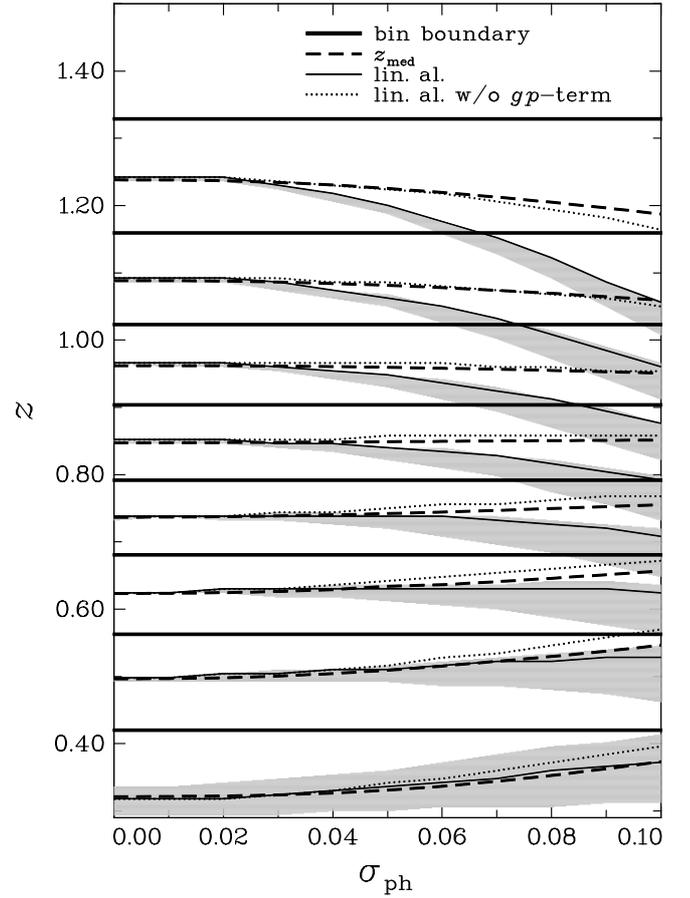}
\caption{Optimal nulling redshift $z_{\rm null}$ as a function of photometric redshift dispersion $\sigma_{\rm ph}$. Plotted are the results for different GI signals, including the linear alignment model with and without $gp$-term, and the power law model with slopes $s=\bc{0.1,0.4,0.7}$. Solid curves correspond to $z_{\rm null}$ for the linear alignment model, evaluated at the central angular frequency bin. Excluding the $gp$-term for this setup results in the dotted line. The gray areas indicate the range of $z_{\rm null}$ for all intrinsic alignment models considered, evaluated at the lowest and highest angular frequency bin each. In addition, the bin boundaries are shown as thick solid lines, while the median redshifts of the redshift probability distributions are represented by thick dashed curves.}
\label{fig:znull_phot}
\end{figure}

In the bottom panel of Fig.$\,$\ref{fig:plotpgi} we plot results for a large redshift uncertainty of $\sigma_{\rm ph}=0.1$. Deviations of the redshift dependence of the GI signal from the best-fitting $g^{(j)}\br{\chi(\hat{z}_i)}$ are visible particularly for the lowest bin considered, i.e. for $j=i+1$, and the bin at the highest redshift. The latter effect can be ascribed to the large width and asymmetry of the corresponding redshift probability distribution, see Fig.$\,$\ref{fig:zdistributions}. The GI power spectrum shifts to higher values for bins $j=i+1$ and $\sigma_{\rm ph} \gg 0$ because of the $gp$-term, which has the strongest contribution for adjacent photometric redshift bins. Accordingly, the GI signal is significantly smaller for bins $j=i+1$ if calculated without the $gp$-term, and a lensing efficiency that fits the GI term much better, i.e. with smaller $R^2(A_P,z_{\rm null})$, can be found. Since $P_{\rm GI}^{(ij)}(\ell)$ without the $gp$-term is generally best-fit by lensing efficiencies with higher $\hat{z}_i$ than the power spectrum with $gp$-term, $R^2$ attains its minimum at higher $\hat{z}_i$, as is also evident from Fig.$\,$\ref{fig:znull_leastsquares}.

We repeat the determination of $z_{\rm null}$ for all relevant initial bins, for the GI power spectrum at the lowest and highest angular frequency bin in addition to the central one, and varying $\sigma_{\rm ph}$, our findings being depicted in Fig.$\,$\ref{fig:znull_phot}. The gray regions cover the range of resulting curves for all four considered GI models (linear alignment; power law with $s=\bc{0.1,0.4,0.7}$), evaluated at the lowest, central, and highest angular frequency bin each. Hence, these regions should mark to good accuracy the possible range of $z_{\rm null}$ for any GI signal. In addition, curves representing the photometric redshift bin boundaries, the median redshifts of the distributions, and $z_{\rm null}$ for the linear alignment model, computed for the central angular frequency bin with and without the $gp$-term are shown.

In the regime of $\sigma_{\rm ph}$ in which nulling performs excellently, i.e. $\sigma_{\rm ph} \lesssim 0.04$ (Fig.$\,$\ref{fig:sigmaphot_total}), we find that the median redshifts are very close to the optimal nulling redshifts. Only for the lowest initial bin the allowed region of $z_{\rm null}$ is broader, but still well-fit by the median redshift. Using the central redshifts $z^{(i)}_{\rm c}$ as nulling redshifts proves to be a fair approximation if the underlying redshift probability distributions are not too asymmetric, as is for instance the case in our model of redshift distributions except for the distributions at the lowest and highest median redshift. These results confirm that variant (C) with nulling at the median redshifts yields indeed the best performance for a survey with small redshift dispersion. As can also be concluded from Fig.$\,$\ref{fig:znull_phot}, variant (B) works only slightly less effectively in this case.

Regarding the behavior of the curves for large $\sigma_{\rm ph}$, $z_{\rm null}$ considerably deviates from its values at small redshift errors, partially crossing the original photometric redshift bin boundaries. While the median redshifts at least qualitatively follow the change in $z_{\rm null}$ with increasing $\sigma_{\rm ph}$ by trend, the $z^{(i)}_{\rm c}$ of nulling variant (B) represent the actual $z_{\rm null}$ even worse, as the results of Fig.$\,$\ref{fig:sigmaphot_total} verify. The drop of $z_{\rm null}$ for the higher initial bins can almost entirely be explained by the $gp$-term contribution. Its removal produces curves that keep close to the median redshifts, see Fig.$\,$\ref{fig:znull_phot}. The remaining offsets of $z_{\rm null}$ from the median redshifts presumably originate from the variation of the integrand in (\ref{eq:limberGI}) across the broad distribution of the initial bins. However, since we compute the GI power spectrum only for single $\ell$-bins, the accuracy in the calculation of $z_{\rm null}$ is limited. This holds true in particular for broad redshift distributions, as the widening of the gray regions, which is dominated by the scatter of the curves computed for different angular frequency bins, indicates.

\section{Influence of further characteristics of the redshift distribution}
\label{sec:furthercharacteristics}

\subsection{Catastrophic outliers}
\label{sec:outliers}

Future cosmic shear data, in particular for space-based surveys incorporating infrared bands \citep{abdalla07}, will be able to rely on exquisite multi-band photometry, so that the fraction of catastrophic failures in the assignment of photometric redshifts will be kept at a very low level. A significant fraction of outliers in the redshift probability distributions would have a devastating effect on the removal of intrinsic alignment. For instance, consider a photometric redshift bin $i$ at relatively high redshift. If it mistakenly contains galaxies whose true redshift is low, these would produce a strong GI signal when correlated with another high redshift background bin $j$.

\begin{figure*}[ht!]
\begin{minipage}[c]{.8\textwidth}
\centering
\includegraphics[scale=.59,angle=270]{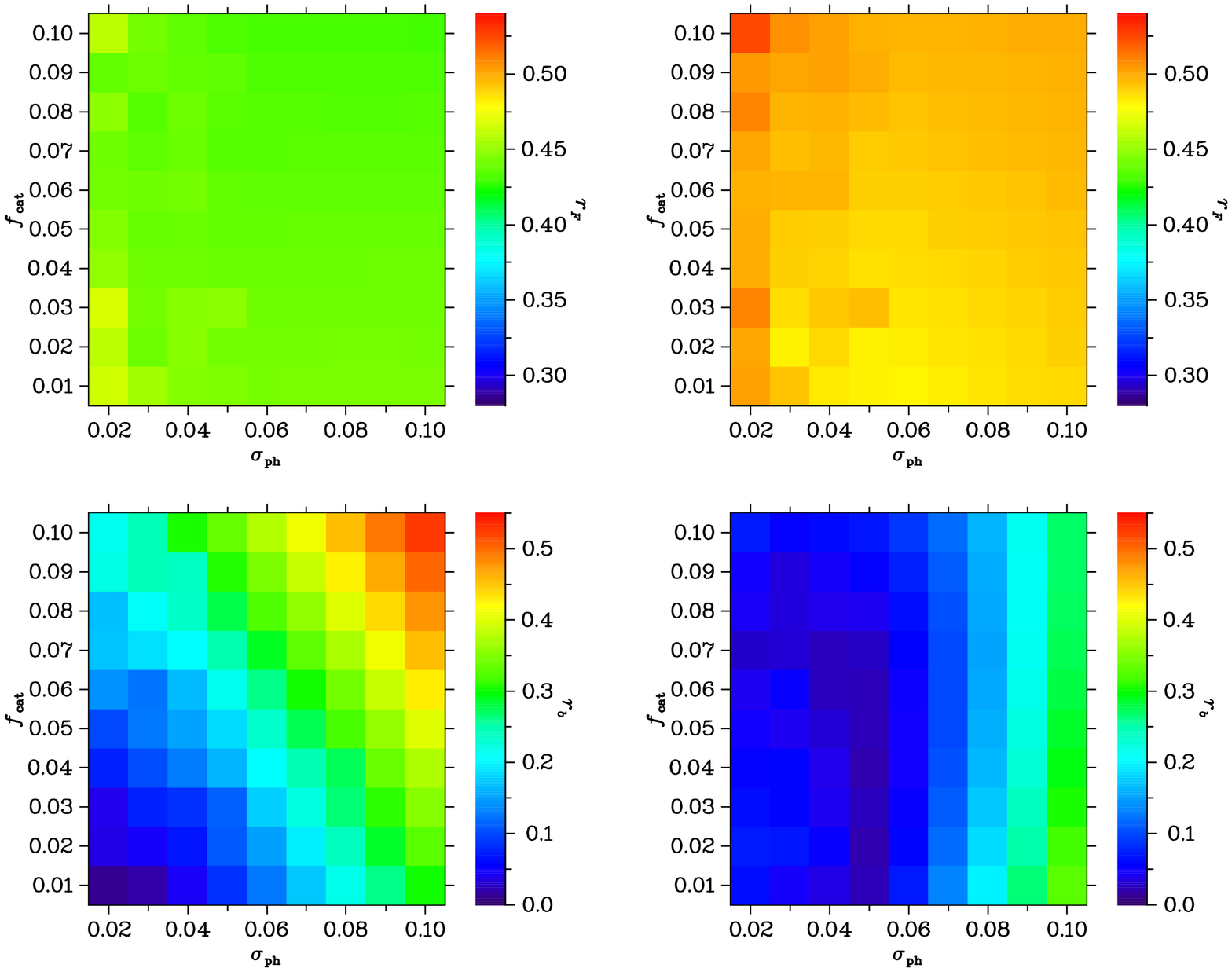}
\includegraphics[scale=.59,angle=270]{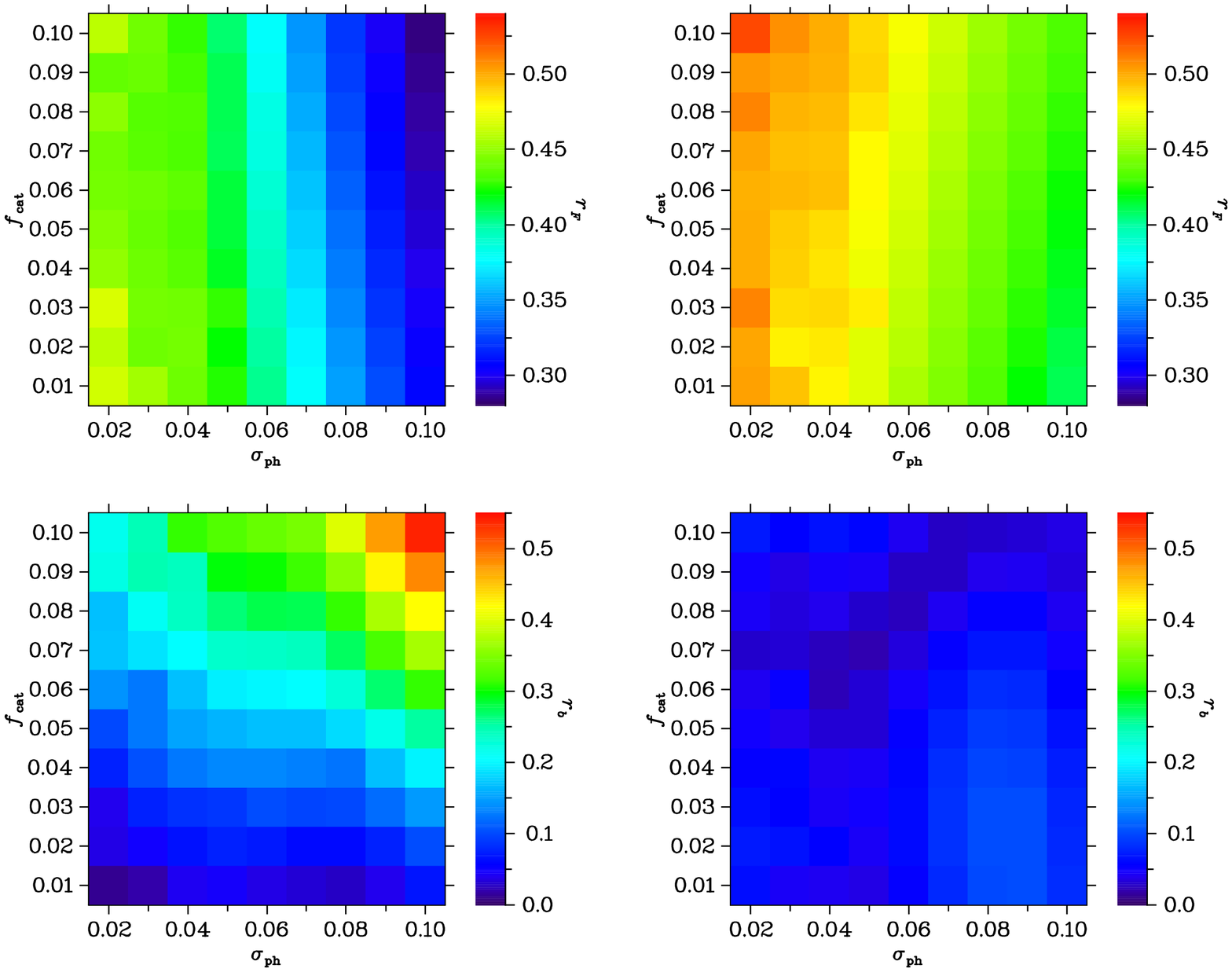}
\end{minipage}%
\begin{minipage}[c]{.2\textwidth}
\caption{Ratios of average statistical and systematic errors $r_F$ and $r_b$ as a function of photometric redshift dispersion $\sigma_{\rm ph}$ and outlier fraction $f_{\rm cat}$. The offset of the outlier distributions has been fixed at $\Delta_z=1$. As systematic the linear intrinsic alignment model, downscaled by a factor of five, has been employed. To obtain the bottom four panels, the calculations were repeated, now including the weighting scheme outlined in Sect.$\,$\ref{sec:controladjacent}. \textit{Left}: Results for nulling which takes into account knowledge of the redshift probability distributions, i.e. variant (C). In panels 1 and 3 $r_F$ is shown, and in panels 2 and 4 $r_b$. \textit{Right}: Same as before, but for nulling with referencing to the centers of the photometric redshift bins, i.e. variant (B). }
\label{fig:bias_fcat}
\end{minipage}
\end{figure*}

We compute the ratios $r_F$ and $r_b$ now as functions of both $\sigma_{\rm ph}$ and $f_{\rm cat}$, keeping the offset fixed at $\Delta_z=1.0$. To judge the effect of outliers, it is important to note that $f_{\rm cat}$ is not the true fraction of catastrophics, but $r_{\rm out}$ as given by Fig.$\,$\ref{fig:realfcat}. Results for $r_F$ and $r_b$ are given in Fig.$\,$\ref{fig:bias_fcat} for the linear intrinsic alignment model as the systematic, again downscaled by a factor of five. The left column shows results for nulling variant (C), the right column for variant (B), where in the bottom four panels the weighting scheme (\ref{eq:gaussianweights}) has been applied in addition.

\begin{figure*}[t]
\centering
\includegraphics[scale=.8,angle=270]{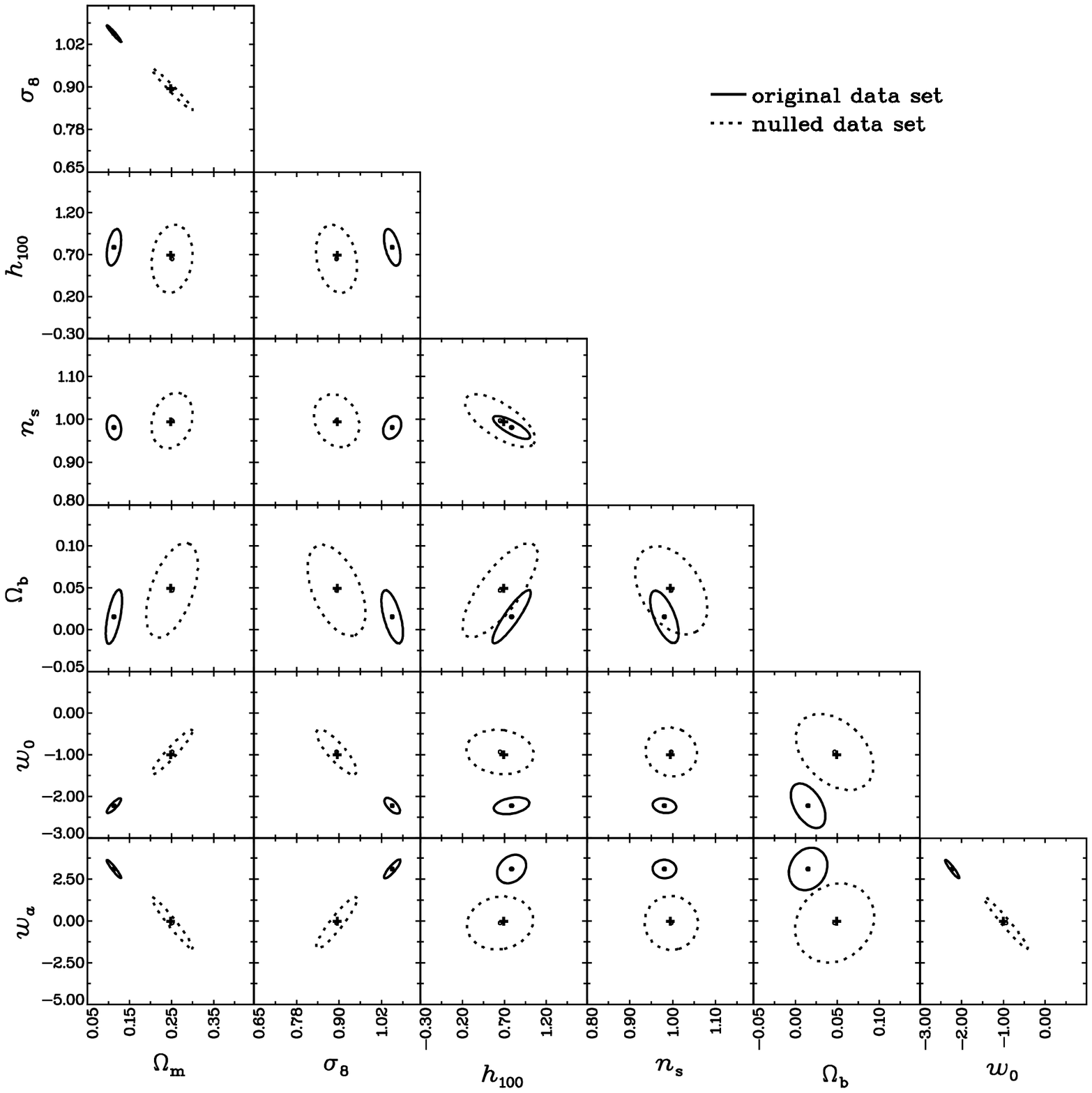}
\caption{Parameter constraints before and after nulling. Shown are the two-dimensional marginalized $2\sigma$-errors for the original data set as solid curves and for the nulled data set as dotted curves. The fiducial parameter values are marked by the crosses. The survey has been divided into $N_z=10$ photometric redshift bins. Photometric redshift errors are characterized by $\sigma_{\rm ph}=0.05$, $f_{\rm cat}=0.05$, and $\Delta_z=1.0$. As systematic the linear alignment model, downscaled by a factor of five, has been employed. The nulling was done using variant (B), including the weighting scheme outlined in Sect.$\,$\ref{sec:controladjacent}.}
\label{fig:bias_plot_fcat}
\end{figure*}

Inspecting the plots obtained without the weighting scheme first, one sees that as before, $r_F$ varies only little with the parameters of photometric redshift, varying around $45\,\%$ for variant (C). Variant (B) retains slightly more information than (C), i.e. around $50\,\%$, which is in accordance with Figs.$\,$\ref{fig:comparenullingeffect} and \ref{fig:sigmaphot_total}. Moreover, the fraction of catastrophic outliers indeed has a strong effect on the ability of nulling to remove the GI systematic. Variant (C) performs well for high quality redshifts, but $r_b$ increases significantly when increasing both $\sigma_{\rm ph}$ and $f_{\rm cat}$, reaching $r_b \approx 0.5$ for $\sigma_{\rm ph} = 0.1$ and $f_{\rm cat} = 0.1$. Contrary to this, variant (B) proves to be much more robust against catastrophic outliers, still reducing the average bias by about a factor of ten for $\sigma_{\rm ph} \leq 0.05$ and any outlier fraction considered here. The performance merely degrades for large $\sigma_{\rm ph}$, but remains below $r_b \approx 0.3$ in the case of the linear alignment model, see also Fig.$\,$\ref{fig:sigmaphot_total}.

\begin{table*}[t]
\begin{minipage}[c]{.18\textwidth}
\caption{Errors on cosmological parameters for three exemplary data sets with different photometric redshift errors. \textit{Top}: Ratios $r_F$ and $r_b$ for the three data sets considered. Moreover, the parameters specifying the photometric redshift errors and the nulling variant used are given. The offset of outliers is fixed at $\Delta_z=1.0$ for all sets. The linear alignment model has been used throughout as systematic, as well as the weighting scheme of Sect.$\,$\ref{sec:controladjacent}. Note that set no.$\,$2 is the underlying data for the results of Fig.$\,$\ref{fig:bias_plot_fcat}. \textit{Bottom}: Marginalized statistical errors $\sigma$, biases $b$, total errors $\sigma_{\rm tot}$, and $b_{\rm rel}$ for every cosmological parameter, shown for both original and nulled data sets. Besides, the ratios of statistical errors and biases before and after nulling are given.}
\end{minipage}%
\begin{minipage}[c]{.82\textwidth}
\centering
\begin{tabular}[t]{c||cccc||cc}
set & $\sigma_{\rm ph}$ & $f_{\rm cat}$ & $r_{\rm out}$ & nulling & $r_F$ & $r_b$\\
\hline
\hline
1 & 0.03 & 0.01 & 0.007 & (C) & 0.438 & 0.026\\
2 & 0.05 & 0.05 & 0.032 & (B) & 0.475 & 0.039\\
3 & 0.07 & 0.10 & 0.060 & (B) & 0.465 & 0.028\\
\end{tabular}
\vspace*{0.7cm}\\
\begin{tabular}[t]{cc||cccc||cccc||cc}
set & par. & \multicolumn{4}{|c||}{original data} & \multicolumn{4}{|c||}{nulled data} & \multicolumn{2}{|c}{ratios}\\
 && $\sigma$ & $b$ & $\sigma_{\rm tot}$ & $b_{\rm rel}$ & $\sigma$ & $b$ & $\sigma_{\rm tot}$ & $b_{\rm rel}$ & $\frac{\sigma_{\rm null}}{\sigma_{\rm orig}}$ & $|\frac{b_{\rm null}}{b_{\rm orig}}|$ \\
\hline
\hline
1 & $\Omega_{\rm m}$ & 0.008 & -0.137 & 0.137 & -16.921 & 0.023 & -0.003 & 0.023 & -0.137 & 2.849 & 0.023\\     
& $\sigma_8$ & 	       0.012 & 0.166 & 0.167 & 14.290 & 0.030 & 0.004 & 0.030 & 0.125 & 2.557 & 0.022    \\     
& $h_{100}$ & 	       0.104 & 0.109 & 0.151 & 1.042 & 0.213 & -0.001 & 0.213 & -0.003 & 2.043 & 0.006  \\     
& $n_{\rm s}$ &        0.014 & -0.012 & 0.018 & -0.882 & 0.036 & -0.001 & 0.036 & -0.029 & 2.615 & 0.086 \\     
& $\Omega_{\rm b}$ &   0.015 & -0.032 & 0.035 & -2.032 & 0.031 & -0.001 & 0.031 & -0.045 & 1.989 & 0.044 \\     
& $w_0$ & 	       0.078 & -1.231 & 1.233 & -15.845 & 0.247 & -0.034 & 0.249 & -0.136 & 3.173 & 0.027\\     
& $w_a$ & 	       0.250 & 3.123 & 3.133 & 12.486 & 0.737 & 0.097 & 0.743 & 0.132 & 2.946 & 0.031    \\     
\hline
2 & $\Omega_{\rm m}$ & 0.009 & -0.136 & 0.136 & -15.674 & 0.025 & 0.003 & 0.025 & 0.140 & 2.830 & 0.025 \\    
& $\sigma_8$ & 	       0.012 & 0.165 & 0.166 & 13.316 & 0.031 & -0.002 & 0.031 & -0.057 & 2.510 & 0.011 \\     
& $h_{100}$ & 	       0.109 & 0.095 & 0.145 & 0.871 & 0.203 & -0.042 & 0.207 & -0.209 & 1.859 & 0.447  \\     
& $n_{\rm s}$ &        0.014 & -0.014 & 0.020 & -0.973 & 0.033 & 0.003 & 0.033 & 0.075 & 2.352 & 0.181  \\     
& $\Omega_{\rm b}$ &   0.016 & -0.034 & 0.038 & -2.101 & 0.030 & -0.002 & 0.030 & -0.084 & 1.831 & 0.073 \\     
& $w_0$ & 	       0.085 & -1.225 & 1.228 & -14.486 & 0.262 & 0.067 & 0.270 & 0.254 & 3.094 & 0.054 \\     
& $w_a$ & 	       0.271 & 3.132 & 3.143 & 11.559 & 0.765 & -0.109 & 0.773 & -0.143 & 2.825 & 0.035 \\     
\hline
3 & $\Omega_{\rm m}$ & 0.010 & -0.135 & 0.135 & -14.090 & 0.026 & -0.002 & 0.026 & -0.075 & 2.758 & 0.015\\          
& $\sigma_8$ & 	       0.014 & 0.164 & 0.164 & 12.066 & 0.033 & 0.005 & 0.034 & 0.145 & 2.466 & 0.030    \\          
& $h_{100}$ & 	       0.116 & 0.079 & 0.140 & 0.676 & 0.218 & -0.042 & 0.222 & -0.194 & 1.879 & 0.538  \\          
& $n_{\rm s}$ &        0.015 & -0.016 & 0.022 & -1.100 & 0.037 & -0.002 & 0.037 & -0.065 & 2.458 & 0.145 \\          
& $\Omega_{\rm b}$ &   0.017 & -0.038 & 0.041 & -2.157 & 0.032 & -0.005 & 0.032 & -0.168 & 1.828 & 0.142 \\          
& $w_0$ & 	       0.095 & -1.211 & 1.215 & -12.773 & 0.283 & 0.021 & 0.284 & 0.073 & 2.986 & 0.017 \\          
& $w_a$ & 	       0.302 & 3.127 & 3.142 & 10.360 & 0.832 & 0.042 & 0.833 & 0.050 & 2.755 & 0.013    \\          
\end{tabular}
\label{tab:errors}
\end{minipage}
\end{table*}

Introducing the weighting scheme for adjacent photometric redshift bins to the nulling technique modifies its performance substantially. For $\sigma_{\rm ph} \lesssim 0.05$ the changes are small, as expected. The larger $\sigma_{\rm ph}$, the more adjacent bin combinations are downweighted, the larger the decrease in $r_F$. The ratio $r_F$ drops by up to 0.15 in the case of variant (C). At the same time the region in which $r_b$ is desirably small extends siginificantly towards larger $\sigma_{\rm ph}$. While this improvement is mostly relevant in the regime of low outlier rates for variant (C), variant (B) achieves $r_b \lesssim 0.1$ across the full range of $\sigma_{\rm ph}$ and $f_{\rm cat}$ considered. In other words, nulling can reduce the GI contamination by at least a factor of 10 for all realistic configurations of redshift errors, given that the GI systematics we consider should be close to a worst case. The even stronger biases caused by the power law models (Fig.$\,$\ref{fig:sigmaphot_total}) are mostly due to the $gp$-term and can thus also be expected to curb down on applying the weighting scheme.

To summarize our findings, we present our different error measures for three exemplary models in Table \ref{tab:errors}. The three sets represent surveys with high (set 1), medium (set 2), and low (set 3) quality redshift information, with parameters $\sigma_{\rm ph}$ and $f_{\rm cat}$ as given in the table. According to the results of the foregoing sections we use variant (C) for the high-quality set 1, and variant (B) for the other configurations, always including the weighting scheme for adjacent photometric redshift bins. For all sets, the survey is divided into $N_z=10$ redshift bins, the downweighted linear alignment model is used as GI signal, and $\Delta_z=1.0$ is fixed. For all these models nulling retains about $45\,\%$ of the statistical power in terms of $r_F$ and depletes the GI contamination by about a factor of 30. Figure \ref{fig:bias_plot_fcat} shows two-dimensional marginalized $2\sigma$-error contours before and after nulling for set 2. Note that since we did not add any priors to the Fisher matrix calculation, negative values for e.g. $\Omega_{\rm b}$ are not excluded.

\subsection{Uncertainty in redshift distribution parameters}
\label{sec:zdisknowledge}

The parameters characterizing the redshift distributions are determined from data, for instance by making use of a spectroscopic subsample of galaxies. Hence, there is also uncertainty in the shape of the $p^{(i)}(z)$, or equivalently, in the parameters describing the redshift distributions such as $z_{\rm med}$, or $\sigma_{\rm ph}$. The performance of variant (C), which explicitly takes into account information about the redshift distributions, will clearly be affected by this uncertainty, as shall be investigated in the following.

\begin{figure}[t]
\centering
\includegraphics[scale=.35,angle=270]{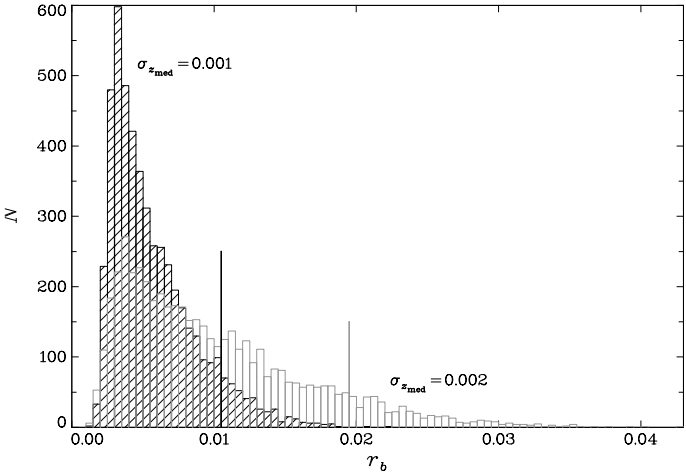}
\caption{Distribution of $r_b$ for 5000 Monte-Carlo samples of the set of $z_{\rm med}$, using a model with $\sigma_{\rm ph}=0.03$ and no catastrophic outliers. The black hatched distribution was obtained for a scatter of $\sigma_{z_{\rm med}}=0.001$, the gray distribution for $\sigma_{z_{\rm med}}=0.002$. The vertical lines mark the limit $\bar{r}_b$, which is chosen such that $r_b<\bar{r}_b$ for $90\,\%$ of all samples.}
\label{fig:montecarlo}
\end{figure}

We quantify the uncertainty in the redshift distributions in terms of the median redshift, allowing for a Gaussian scatter with width $\sigma_{z_{\rm med}}$ around the true value of $z_{\rm med}$ for every redshift bin. Then Monte-Carlo samples of sets of $z_{\rm med}$ are drawn from these distributions and used to subsequently compute nulling weights, do the Fisher analysis of the nulled data set, and obtain the ratio $r_b$. As input we use a set of power spectra calculated for $N_z=10$ bins with $\sigma_{\rm ph}=0.03$ and without catastrophic outliers. For high-quality redshift information that nulling variant (C) is suited for one can adopt the requirements on $\sigma_{z_{\rm med}}$ of planned satellite missions like Euclid, targeting $\sigma_{z_{\rm med}}=0.001$ and demanding at least $\sigma_{z_{\rm med}}=0.002$. Drawing 5000 Monte-Carlo samples each for both of these values of $\sigma_{z_{\rm med}}$ produces the distributions of $r_b$ displayed in Fig.$\,$\ref{fig:montecarlo}.

For each histogram a value $\bar{r}_b$ is marked, defined such that $r_b<\bar{r}_b$ for $90\,\%$ of all samples. We find $\bar{r}_b \approx 0.010$ for $\sigma_{z_{\rm med}}=0.001$ and $\bar{r}_b \approx 0.019$ for $\sigma_{z_{\rm med}}=0.002$. The distributions peak at the value $r_b \approx 0.003$, which results from using the $z_{\rm med}$ as nulling redshifts (see Fig.$\,$\ref{fig:sigmaphot_total}). Given a non-vanishing photometric redshift error, $z_{\rm med}$ is not necessarily the optimal choice, and indeed samples with $r_b < 0.003$ exist, although the histograms decline rapidly for small $r_b$. The distribution for $\sigma_{z_{\rm med}}=0.002$ is much shallower and decreases only slowly for $r_b > 0.003$, resulting in a $\bar{r}_b$ about twice as big as for $\sigma_{z_{\rm med}}=0.001$. Hence, nulling variant (C) requires knowledge of the form of the redshift distribution comparable to the planned goals of future satellite missions to fully demonstrate its potential. Any moderate deviation of the nulling redshifts from its optimum, approximated by the $z_{\rm med}$, results in a significant increase in residual bias.

On the other hand, nulling variant (B) does not rely on detailed knowledge about the $p^{(i)}(z)$ and performed well over a wide range of redshift distribution characteristics, but only when including the Gaussian weighting scheme of adjacent redshift bins. The latter procedure does depend on the form of the redshift distributions to a certain extent as the width of the weight should be chosen such that the Gaussian covers the range of overlap between the redshift distributions, which in turn depends on $\sigma_{\rm ph}$. However, general information about the width of redshift distribution is mandatory for all upcoming cosmic shear surveys. Since the width of the Gaussian in (\ref{eq:gaussianweights}) can in principle be chosen arbitrarily, one can always adjust this width to safely suppress the $gp$-term.

\section{Summary \& conclusions}
\label{sec:conclusions}

In this paper we investigated the performance of the nulling technique as proposed by JS08, designed to geometrically eliminate the contamination by gravitational shear-intrinsic ellipticty correlations. In the presence of realistic photometric redshift information and errors we considered both the information loss due to nulling and the amount of residual bias. We suggested several modifications and improvements to the original technique, which we summarize by providing a recipe on how to apply nulling to a cosmic shear tomography data set.

(1) Decide on which variant of nulling is best suited for the data set. If the data has precise information about the redshift distributions, and if these distributions have a small scatter and negligible outlier fraction, then variant (C), which takes into account this information, should be chosen. Otherwise variant (B) is preferable, if combined with a Gaussian downweighting of combinations of adjacent photometric redshift bins. This weighting scheme is necessary since overlapping redshift distributions can cause a swap of foreground and background galaxies, which produces a GI signal that cannot be controlled by means of nulling. Both variants perform considerably better than the original referencing suggested by JS08.

(2) Calculate the nulling weights, depending on the variant chosen. This work defines these weights such that nulling can be interpreted as an orthonormal transformation of the cosmic shear data vector. Since the weights are composed of comoving distances, one has to assume a cosmology to compute them. An incorrect choice of parameters affects the GI removal and could in principle cause an even stronger bias on parameter estimates. We showed that any reasonable choice of cosmological parameters will produce equally suited nulling weights -- one could even start with the resulting, largely biased parameters of the analysis of the original data set. Iteratively using the parameter estimates as input for a renewed nulling analysis renders the final results independent of any initial assumptions.

(3) Compute nulled cosmic shear measures from the nulling weights and the tomography measures available. As nulling does not depend on angular scales, any measure such as the shear correlation functions or the aperture mass dispersion are suited. The number and size of photometric redshift bins should be chosen such that the overlap of the corresponding redshift distributions is kept at a minimum. Although nulling reduces the GI signal also for a division into 5 bins, we found that $N_z \geq 10$ is required to achieve good performance. Auto-correlations should be excluded from the analysis because of the potential contamination by an II signal. Applying the Gaussian weighting scheme will also reduce the II contamination in shear measures of adjacent photometric redshift bins.

Performing a likelihood analysis with the nulled data should then yield parameter constraints that have a low residual bias due to intrinsic alignment contributions. However, we outlined that nulling inevitably reduces the information content in the data, even if spectroscopic redshifts were available. We demonstrated that lensing information, integrated over wide redshift ranges, is eliminated together with the GI term, which can finally be traced back to the distinct, but still similar dependence on redshift of the lensing and GI signal. In terms of our figure of merit $r_F$ we found that of the order $50\,\%$ of the statistical power is lost. The loss decreases for larger $N_z$, so that in contrast to a lensing-only analysis $N_z \gg 5$ is desirable, which is in accordance with earlier work \citep[JS08]{bridle07}.

In this paper we have not exploited any feature of intrinsic alignments apart from its dependence on redshift. However, observations suggest that the strongest intrinsic alignment signal stems from luminous galaxies \citep{mandelbaum06,hirata07}. Photometric redshift estimates for these bright galaxies usually have a much smaller scatter \citep{ilbert08}, so that nulling may work better on this important subset. Thus, our conclusions on the performance of the nulling technique should be conservative. 

Given excellent redshift information, nulling variant (C) reduces the bias, averaged over all parameters considered as defined in (\ref{eq:averagebias}), by at least a factor of 100. To achieve this goal, stringent conditions like $\sigma_{\rm ph} \lesssim 0.03$, a negligible fraction of catastrophic outliers, and an uncertainty in the median redshift $\sigma_{z_{\rm med}} \lesssim 0.001$ hold. Even future space-based surveys will fulfill these requirements only for a brighter subsample of galaxies (which are expected to have the strongest intrinsic alignment signal though), but still this nulling version could serve as a valuable consistency check. To suppress the GI signal by a factor of about 20, the conditions are moderately released, in particular on $\sigma_{\rm ph}$, in case the Gaussian weighting is used. Moreover, we determined optimal nulling redshifts, demonstrating that for accurate redshift information variant (C) is close to the best configuration possible in this geometric approach.

Throughout the considered parameter plane, spanned by $f_{\rm cat} \leq 0.1$ (corresponding to a true outlier fraction of $\leq 6\,\%$) and $\sigma_{\rm ph} \leq 0.1$, the nulling version based on variant (B) was capable of reducing the average bias by at least a factor of 10. Consequently, the requirements on photometric redshift parameters are low in this case. Merely a number $N_z \geq 10$ of photometric redshift bins, for which the width of the underlying redshift distributions should be known, is demanded -- readily achieved by the majority of future cosmic shear surveys. Although we showed that the functional behavior of the residual bias is similar for all considered models, the values of the residual bias depend on the actual form of the GI signal. Since all models considered in this work produce severe parameter biases, we have further reason to believe that the numbers for the performance of the nulling technique given above should be understood as conservative.

We have neglected the contamination by the II signal in all our considerations, arguing that the nulling could be preceded by an appropriate II removal technique. While for disjoint photometric redshift bins the II signal does not appear in the transformed data at all, it was demonstrated that, for realistic situations, ignoring the II term may cause a significant contamination of a subset of the nulled power spectra. On the other hand, this restriction of the II signal to certain nulled power spectra only could also allow for a removal of II after nulling. In any case, the ultimate goal is a combined geometrical treatment of all intrinsic alignment contributions, which is subject to forthcoming work.

Although we sampled only a fraction of the huge parameter space spanned by the various photometric redshift parameters, GI models, and nulling variants, it should be possible to draw a wide range of conclusions from this work. For instance, a relevant question is how a cosmic shear data set should be binned in order to remove intrinsic alignment and keep a maximum of information. The bin boundaries should be chosen such that the overlap of the corresponding redshift distributions is minimal, as long as the distributions do not become too asymmetric. Re-inspecting Fig.$\,$\ref{fig:nzbias}, the number of bins should be as big as the photometric redshift scatter allows, i.e. the width of the bins should not become smaller than about $\sigma_{\rm ph}(1+z)$ since otherwise no more information is added. As our results show, the photometric redshift scatter does not necessarily limit the level to which the GI signal can be eliminated, but then it places strong bounds on the remaining power to constrain cosmological parameters in the nulled data set, see Fig.$\,$\ref{fig:bias_fcat}.

We emphasize that, in spite of defining GI signals to quantify the bias removal, the nulling technique itself does not rely on any information about intrinsic alignment except for the well-known redshift dependence of the GI term. In principle, nulling is also applicable to data sets in which the GI contribution dominates over lensing. Provided a sufficient suppression, it would be possible to recover the cosmic shear signal by nulling the data. Besides, nulling is not restricted to cosmic shear at the two-point level. Concerning three-point statistics, gravitational shear-intrinsic ellipticity cross terms, GII and GGI, may constitute an even more serious contamination \citep{semboloni08}. The geometric principle of nulling can be applied to tomography bispectra and related real-space measures in a straightforward manner (Shi et al., in preparation).

Due to the significant information loss of nulling, this technique is most probably not desirable as the standard GI removal tool for future surveys, so that the need for both an improved understanding of intrinsic alignment and high-performance removal techniques that take knowledge about the GI models into account persists. Still, with its very low level of input assumptions, nulling serves as a valuable cross-check for these model-dependent techniques yet to be developed and as such can contribute to the credibility of cosmic shear as a powerful and robust cosmological probe.

\begin{acknowledgements}
We would like to thank our referee for a very helpful report. BJ acknowledges support by the Deutsche Telekom Stiftung and the Bonn-Cologne Graduate School of Physics and Astronomy. This work was supported by the Priority Programme 1177 of the Deutsche Forschungsgemeinschaft, by the Transregional Collaborative Research Centre TR 33 of the DFG, and the RTN-Network DUEL of the European Commission.
\end{acknowledgements}

\bibliographystyle{aa}

\begin{appendix}
\section{Fisher matrix for a parameter-dependent data vector}
\label{sec:appendix}

In the following we explicitly calculate the Fisher matrix for a data vector $\vek{y}$, transformed according to (\ref{eq:datatrafo}), where the transformation $\vek{T}$ depends on the parameters to be determined. We closely follow the derivation of the Fisher matrix presented in \citet{tegmark97}. A comma notation is used to indicate derivatives with respect to parameters.

For $\vek{y}$ the Gaussian log-likelihood reads
\eqa{
\nn
- \ln L_y(\vek{y}|\vek{p}) &=& \frac{N_{\rm d}}{2} \ln 2\pi + \frac{1}{2} \ln \det C_y\\
\label{eq:loglikey}
&& \hspace*{.5cm} + \frac{1}{2} \bb{\vek{y}-\bar{\vek{y}}}^\tau C_y^{-1} \bb{\vek{y}-\bar{\vek{y}}}\;,
}
where we dropped the arguments of $\vek{y}$ and $C_y$ for notational convenience. Again, the expectation value of a data vector is indicated by a bar over the corresponding variable name. Making use of the matrix identity $\ln \det C = {\rm tr} \ln C$, and defining the matrix $D_y \equiv \br{\vek{y}-\bar{\vek{y}}} \br{\vek{y}-\bar{\vek{y}}}^\tau$, one arrives at
\eq{
\label{eq:loglikewithtrace}
- \ln L_y(\vek{y}|\vek{p}) = \frac{N_{\rm d}}{2} \ln 2\pi + \frac{1}{2}\; {\rm tr} \bc{\ln C_y + C_y^{-1} D_y}\;.
}
According to the derivation in \citet{tegmark97}, the second derivative of (\ref{eq:loglikewithtrace}) reads\footnote{Note that there is a typo in Eq.$\,$(14) of \citet{tegmark97}: A factor $C^{-1}$ should be eliminated from the last term.}
\eqa{
\nn
- \bc{ \ln L_y(\vek{y}|\vek{p}) }_{,\mu\nu} &=& \frac{1}{2}\; {\rm tr}\; \Bigl\{ C_y^{-1} {C_y}_{,\mu\nu} - C_y^{-1} {C_y}_{,\mu\nu} C_y^{-1} D_y\\ \nn
&& + C_y^{-1} {C_y}_{,\nu} C_y^{-1} {C_y}_{,\mu} C_y^{-1} D_y - C_y^{-1} {C_y}_{,\mu} C_y^{-1} {D_y}_{,\nu}\\ 
\label{eq:2ndderivative}
&& - C_y^{-1} {C_y}_{,\nu} C_y^{-1} {D_y}_{,\mu} + C_y^{-1} {D_y}_{,\mu\nu} \Bigr\}\;,
}
where the rules $\br{\ln C}_{,\mu} = C^{-1} C_{,\mu}$ and $(C^{-1})_{,\mu} = - C^{-1} C_{,\mu} C^{-1}$ were applied. The expectation value of (\ref{eq:2ndderivative}) yields the Fisher matrix, see the definition in (\ref{eq:fisherdef}). We compute the matrix $D_y$ and its derivatives in terms of the original data set,
\eqa{
\label{eq:Dy}
D_y &=& \vek{T} D_x \vek{T}^\tau\;;\\ \nn
{D_y}_{,\mu} &=& \vek{T}_{,\mu} D_x \vek{T}^\tau \! + \! \vek{T} D_x \vek{T}^\tau_{,\mu} \! - \! \vek{T} \bar{\vek{x}}_{,\mu} \br{\vek{x}-\bar{\vek{x}}}^\tau \vek{T}^\tau \! - \! \vek{T} \br{\vek{x}-\bar{\vek{x}}} \bar{\vek{x}}_{,\mu}^\tau \vek{T}^\tau \;;\\ \nn
{D_y}_{,\mu\nu} &=& \vek{T}_{,\mu\nu} D_x \vek{T}^\tau - \br{ \vek{T}_{,\mu} \bar{\vek{x}}_{,\nu} + \vek{T}_{,\nu} \bar{\vek{x}}_{,\mu} + \vek{T} \bar{\vek{x}}_{,\mu\nu}} \br{\vek{x}-\bar{\vek{x}}}^\tau \vek{T}^\tau\\ \nn
&& + \vek{T} D_x \vek{T}^\tau_{,\mu\nu} - \vek{T} \br{\vek{x}-\bar{\vek{x}}} \br{ \vek{T}_{,\mu} \bar{\vek{x}}_{,\nu} + \vek{T}_{,\nu} \bar{\vek{x}}_{,\mu} + \vek{T} \bar{\vek{x}}_{,\mu\nu}}^\tau\\ \nn
&& + \vek{T}_{,\mu} D_x \vek{T}^\tau_{,\nu} - \vek{T} \bar{\vek{x}}_{,\mu} \br{\vek{x}-\bar{\vek{x}}}^\tau  \vek{T}^\tau_{,\nu} - \vek{T}_{,\mu} \br{\vek{x}-\bar{\vek{x}}} \bar{\vek{x}}_{,\nu}^\tau \vek{T}^\tau\\ \nn
&& + \vek{T} \bar{\vek{x}}_{,\mu} \bar{\vek{x}}_{,\nu}^\tau \vek{T}^\tau + \vek{T}_{,\nu} D_x \vek{T}^\tau_{,\mu}\\ \nn
&& - \vek{T} \bar{\vek{x}}_{,\nu} \br{\vek{x}-\bar{\vek{x}}}^\tau  \vek{T}^\tau_{,\mu} - \vek{T}_{,\nu} \br{\vek{x}-\bar{\vek{x}}} \bar{\vek{x}}_{,\mu}^\tau \vek{T}^\tau + \vek{T} \bar{\vek{x}}_{,\nu} \bar{\vek{x}}_{,\mu}^\tau \vek{T}^\tau\;,
}
where $D_x$ is defined in analogy to $D_y$. Using $\ba{\vek{x}} = \bar{\vek{x}}$ and $\ba{\vek{x} \vek{x}^\tau} = C_x +  \bar{\vek{x}}  \bar{\vek{x}}^\tau$, we obtain the expectation values of the former quantities,
\eqa{
\label{eq:Dyexpt}
\ba{ D_y } &=& \vek{T} C_x \vek{T}^\tau = C_y\;;\\ \nn
\ba{ {D_y}_{,\mu} } &=& \vek{T}_{,\mu} C_x \vek{T}^\tau + \vek{T} C_x \vek{T}^\tau_{,\mu}\;;\\ \nn
\ba{ {D_y}_{,\mu\nu} } &=& \vek{T}_{,\mu\nu} C_x \vek{T}^\tau + \vek{T} C_x \vek{T}^\tau_{,\mu\nu} + \vek{T}_{,\mu} C_x \vek{T}^\tau_{,\nu} + \vek{T}_{,\nu} C_x \vek{T}^\tau_{,\mu} \\ \nn
&& \hspace*{.5cm} + \vek{T} \br{ \bar{\vek{x}}_{,\mu} \bar{\vek{x}}_{,\nu}^\tau + \bar{\vek{x}}_{,\nu} \bar{\vek{x}}_{,\mu}^\tau } \vek{T}^\tau\;.
}
With these expressions at hand we calculate the expectation value of (\ref{eq:2ndderivative}),
\eqa{
\label{eq:fishery}
F_{\mu\nu}^y &=& \ba{ - \bc{ \ln L_y(\vek{y}|\vek{p}) }_{,\mu\nu} }\\ \nn
&=& \frac{1}{2}\; {\rm tr}\; \Bigl\{ C_y^{-1} \br{ \vek{T}_{,\nu} C_x \vek{T}^\tau + \vek{T} {C_x}_{,\nu} \vek{T}^\tau + \vek{T} C_x \vek{T}^\tau_{,\nu} }\\ \nn
&& \hspace*{2cm} \times\; C_y^{-1} \br{ \vek{T}_{,\mu} C_x \vek{T}^\tau + \vek{T} {C_x}_{,\mu} \vek{T}^\tau + \vek{T} C_x \vek{T}^\tau_{,\mu} }\\ \nn
&& \hspace*{-.9cm} -  C_y^{-1} \br{ \vek{T}_{,\nu} C_x \vek{T}^\tau + \vek{T} {C_x}_{,\nu} \vek{T}^\tau + \vek{T} C_x \vek{T}^\tau_{,\nu} } C_y^{-1} \br{ \vek{T}_{,\mu} C_x \vek{T}^\tau + \vek{T} C_x \vek{T}^\tau_{,\mu} }\\ \nn
&& \hspace*{-.9cm} -  C_y^{-1} \br{ \vek{T}_{,\mu} C_x \vek{T}^\tau + \vek{T} {C_x}_{,\mu} \vek{T}^\tau + \vek{T} C_x \vek{T}^\tau_{,\mu} } C_y^{-1} \br{ \vek{T}_{,\nu} C_x \vek{T}^\tau + \vek{T} C_x \vek{T}^\tau_{,\nu} }\\ \nn
&& \hspace*{-.9cm} + C_y^{-1} \bigl( \vek{T}_{,\mu\nu} C_x \vek{T}^\tau + \vek{T} C_x \vek{T}^\tau_{,\mu\nu} + \vek{T}_{,\mu} C_x \vek{T}^\tau_{,\nu}\\ \nn
&& \hspace*{2cm} + \vek{T}_{,\nu} C_x \vek{T}^\tau_{,\mu} + \vek{T} \br{ \bar{\vek{x}}_{,\mu} \bar{\vek{x}}_{,\nu}^\tau + \bar{\vek{x}}_{,\nu} \bar{\vek{x}}_{,\mu}^\tau } \vek{T}^\tau \bigr) \Bigr\}\;.
}
Note that the first two terms in (\ref{eq:2ndderivative}) cancel due to $\ba{ D_y } = C_y$. We now make extensive use of the fact that the trace is invariant under cyclic permutations of matrices. Then one readily finds that many terms in the first three lines of (\ref{eq:fishery}) cancel. Expanding $C_y^{-1} =  {\vek{T}^\tau}^{-1} C_x^{-1} \vek{T}^{-1}$, more terms cancel, either directly or after cyclic permutation. This way (\ref{eq:fishery}) reduces to
\eqa{
\label{eq:fishery2}
F_{\mu\nu}^y &=& \frac{1}{2}\; {\rm tr}\; \Bigl\{ C_x^{-1} {C_x}_{,\nu} C_x^{-1} {C_x}_{,\mu} + C_x^{-1} \br{ \bar{\vek{x}}_{,\mu} \bar{\vek{x}}_{,\nu}^\tau + \bar{\vek{x}}_{,\nu} \bar{\vek{x}}_{,\mu}^\tau }\\ \nn
&& + \vek{T}^{-1} \vek{T}_{,\mu\nu} + \vek{T}^\tau_{,\mu\nu} {\vek{T}^\tau}^{-1} - \vek{T}^{-1} \vek{T}_{,\nu} \vek{T}^{-1} \vek{T}_{,\mu} - {\vek{T}^\tau}^{-1} \vek{T}^\tau_{,\nu} {\vek{T}^\tau}^{-1} \vek{T}^\tau_{,\mu} \Bigr\}\;.
}
The first two terms of this expression correspond to the Fisher matrix $F_{\mu\nu}^x$ of the data vector $\vek{x}$, see (\ref{eq:gaussianfisher}). Finally, by employing in addition that ${\rm tr}\, C^\tau = {\rm tr}\, C$ and ${(C^\tau)}^{-1} = {(C^{-1})}^\tau$, one arrives at
\eq{
\label{eq:fishery3}
F_{\mu\nu}^y = F_{\mu\nu}^x + {\rm tr} \bc{\ln \vek{T}}_{,\mu\nu}\;.
}
If we apply the condition $\det \vek{T} = 1$, as required in Sect.$\,$\ref{sec:fisher}, we find ${\rm tr} \ln \vek{T} = \ln \det \vek{T} = 0$, and hence, the Fisher matrices of the original data vector $\vek{x}$ and the transformed one $\vek{y}$ are equivalent. This result is in agreement with (\ref{eq:likey}), which, when transformed to log-likelihood, reads
\eqa{
\label{eq:loglikey2}
- \ln L_y(\vek{y}|\vek{p}) &=& \ln \det \vek{T}(\vek{p})\; - \ln L_x(\vek{x}|\vek{p})\\ \nn
&=& {\rm tr} \bc{ \ln \vek{T}(\vek{p})}\; - \ln L_x(\vek{x}|\vek{p})
}
and reproduces (\ref{eq:fishery3}) after taking derivatives and expectation value. Employing the further simplification that the original covariance $C_x$ does not depend on the parameters, the Fisher matrix can be written as
\eqa{
\label{eq:fishercconst}
F_{\mu\nu} &=& \frac{1}{2}\; {\rm tr} \bc{ C_x^{-1} \br{ \bar{\vek{x}}_{,\mu} \bar{\vek{x}}_{,\nu}^\tau + \bar{\vek{x}}_{,\nu} \bar{\vek{x}}_{,\mu}^\tau }}\\ \nn
&=& \frac{1}{2}\; {\rm tr} \bc{ C_y^{-1} \vek{T} \br{ \bar{\vek{x}}_{,\mu} \bar{\vek{x}}_{,\nu}^\tau + \bar{\vek{x}}_{,\nu} \bar{\vek{x}}_{,\mu}^\tau } \vek{T}^\tau }\;,
}
which, after converting the trace to a sum, yields (\ref{eq:fisherfinal}).

\section{Validity of the bias formalism}
\label{sec:validitybias}

As is evident from Sect.$\,$\ref{sec:GIsignal}, a GI systematic that fits within the error bounds of current observations can attain values of similar order of magnitude as the lensing power spectrum. Besides, due to the similar dependence on geometry, see (\ref{eq:limber}) and (\ref{eq:limberGI}), the effect of adding a GI systematic acts similarly to a change of cosmological parameters, in particular those determining the amplitude of the lensing power spectrum. Consequently, we expect the systematic to produce a strong bias, possibly much larger than the statistical error bounds. While this does not hamper the performance of the nulling technique, it may render the bias formalism as given by (\ref{eq:bias}) invalid. In the following we are going to derive the parameter bias from the log-likelihood, taking special care of approximations and the resulting limitations.

Since we keep the assumption that the signal covariance $C_P$ does not depend on the parameters to be determined, the calculations can be directly done in terms of the $\chi^2$, which is then twice the log-likelihood. For a similar approach see e.g. \citet{taburet09}. We define a fiducial data vector $\vek{P^{\rm f}}$, i.e. the signal in absence of systematic effects, and assume this signal to be contaminated by a systematic $\vek{P^{\rm sys}}$. A set of models $\vek{P}(\vek{p})$, depending on a set of parameters $\vek{p}$, is fitted to the signal, where $\vek{p^{\rm f}}$ denotes the fiducial set of parameters such that $\vek{P}(\vek{p^{\rm f}})=\vek{P^{\rm f}}$. Then the $\chi^2$ reads
\eq{
\label{eq:biasedchisquare}
\chi^2(\vek{p}) = \sum_{\alpha,\beta} \br{P_\alpha(\vek{p}) - P_\alpha^{\rm tot}} \br{C_P^{-1}}_{\alpha\beta} \br{P_\beta(\vek{p}) - P_\beta^{\rm tot}}\;,
}
where $P_\alpha^{\rm tot} \equiv P_\alpha^{\rm f} + P_\alpha^{\rm sys}$. Writing the unbiased $\chi^2$ as
\eq{
\label{eq:unbiasedchisquare}
\chi_0^2(\vek{p}) = \sum_{\alpha,\beta} \br{P_\alpha(\vek{p}) - P_\alpha^{\rm f}} \br{C_P^{-1}}_{\alpha\beta} \br{P_\beta(\vek{p}) - P_\beta^{\rm f}}\;,
}
one can expand (\ref{eq:biasedchisquare}) to yield
\eq{
\label{eq:expandedchisquare}
\chi^2(\vek{p}) = \chi^2(\vek{p^{\rm f}}) + \chi_0^2(\vek{p}) - 2\,\sum_{\alpha,\beta} P_\alpha^{\rm sys} \br{C_P^{-1}}_{\alpha\beta} \br{P_\beta(\vek{p}) - P_\beta^{\rm f}}\;,
}
where $\vek{p^{\rm f}}$ produces the maximum likelihood (or minimum $\chi^2$) in absence of a systematic. Since $\vek{P}(\vek{p^{\rm f}})=\vek{P^{\rm f}}$, $\chi^2(\vek{p^{\rm f}})$ contains only the systematic power spectrum and causes an irrelevant overall rescaling of the $\chi^2$ in parameter space. Hence, the modification of the $\chi^2$ due to the systematic is contained in the last term of (\ref{eq:expandedchisquare}). It can shift the point of maximum likelihood and deform the likelihood in its vicinity, depending on both the parameters and the form of the systematic.

Considering (\ref{eq:biasedchisquare}) again, $\chi^2(\vek{p})$ can be written as a Taylor expansion around the fiducial set of parameters,
\eqa{
\nn
\chi^2(\vek{p}) &=& \chi^2(\vek{p^{\rm f}}) + \sum_i \left. \frac{\partial \chi^2}{\partial p_i} \right|_{\rm f} \br{p_i - p_i^{\rm f}}\\
\label{eq:taylorchisquare}
&& + \frac{1}{2}\; \sum_{i,j} \br{p_i - p_i^{\rm f}} \left. \frac{\partial^2 \chi}{\partial p_i\; \partial p_j} \right|_{\rm f} \br{p_j - p_j^{\rm f}} + {\cal O}\br{p^3}\;,
}
where the subscript ${\rm f}$ indicates that the derivatives are evaluated at $\vek{p^{\rm f}}$. Making again use of $\vek{P}(\vek{p^{\rm f}})=\vek{P^{\rm f}}$, one obtains for the derivatives from (\ref{eq:expandedchisquare})
\eqa{
\label{eq:chisquarederivatives1}
\left. \frac{\partial \chi^2}{\partial p_i} \right|_{\rm f} &=& -2\,\sum_{\alpha,\beta} P_\alpha^{\rm sys} \br{C_P^{-1}}_{\alpha\beta} \left. \frac{\partial P_\beta}{\partial p_i} \right|_{\rm f}\;;\\ 
\nn
\left. \frac{\partial^2 \chi^2}{\partial p_i\; \partial p_j} \right|_{\rm f} &=& 2\,\sum_{\alpha,\beta} \Biggl\{ \left. \frac{\partial P_\alpha}{\partial p_i} \right|_{\rm f}  \br{C_P^{-1}}_{\alpha\beta} \left. \frac{\partial P_\beta}{\partial p_j} \right|_{\rm f}\\ 
\label{eq:chisquarederivatives2}
&& \hspace*{1.5cm} - P_\alpha^{\rm sys} \br{C_P^{-1}}_{\alpha\beta} \left. \frac{\partial^2 P_\beta}{\partial p_i\; \partial p_j} \right|_{\rm f} \Biggr\}\;.
}
Dividing (\ref{eq:chisquarederivatives2}) by 2 yields the Fisher matrix, so that in the case of a biased $\chi^2$ one can define an equivalent to the Fisher matrix as
\eq{
\label{biasedfisher}
{F'}_{\mu\nu} \equiv F_{\mu\nu} - \sum_{\alpha,\beta} P_\alpha^{\rm sys} \br{C_P^{-1}}_{\alpha\beta} \left. \frac{\partial^2 P_\beta}{\partial p_i\; \partial p_j} \right|_{\rm f}\;.
}

\begin{figure}[t]
\centering
\includegraphics[scale=.6]{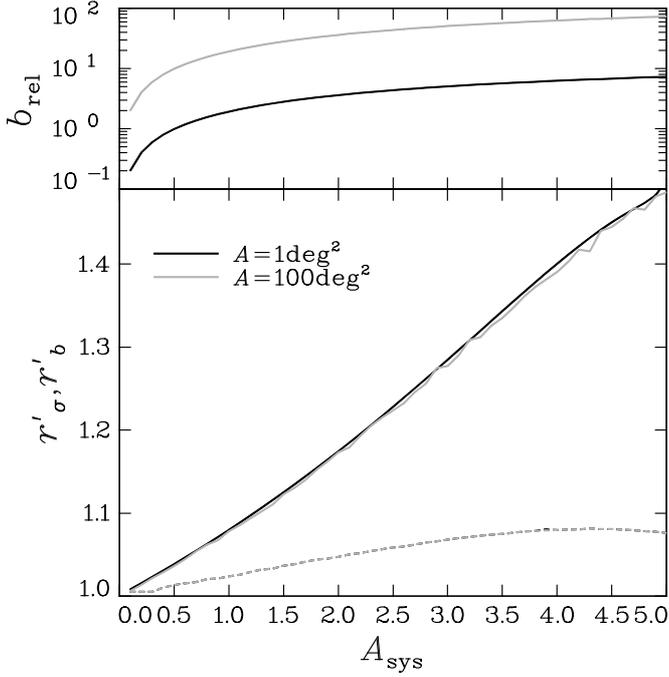}
\caption{Comparison of statistical errors and biases obtained by Fisher matrix and $\chi^2$ calculations. \textit{Top panel}: Ratio of bias over statistical error $b_{\rm rel}$ as a function of the scaling of the systematic $A_{\rm sys}$. Results for a $1\,{\rm deg}^2$ survey are shown as black curves, and for a $100\,{\rm deg}^2$ survey as gray curves. \textit{Bottom panel}: Ratios of the statistical errors ${r'}_\sigma$ and biases ${r'}_b$ as a function of the scaling of the systematic $A_{\rm sys}$. Solid lines correspond to ${r'}_\sigma$, dashed lines to ${r'}_b$. As above, results for a $1\,{\rm deg}^2$ and a $100\,{\rm deg}^2$ survey are shown as black and gray curves, respectively. Note that the curves for ${r'}_b$ almost completely overlap. }
\label{fig:fishlikecomp}
\end{figure}

We want to determine the bias $\vek{b} \equiv \vek{p^{\rm b}} - \vek{p^{\rm f}}$, where $\vek{p^{\rm b}}$ is the point in parameter space where the biased $\chi^2$ attains its minimum. The biased parameter set $\vek{p^{\rm b}}$ is computed from (\ref{eq:taylorchisquare}), using the expansion up to second order, which results in
\eq{
\label{eq:chisquareextremum}
\left. \frac{\partial \chi^2}{\partial p_k} \right|_{\rm b} = -2\,\sum_{\alpha,\beta} P_\alpha^{\rm sys} \br{C_P^{-1}}_{\alpha\beta} \left. \frac{\partial P_\beta}{\partial p_k} \right|_{\rm f} + 2\,\sum_i {F'}_{ki} b_i = 0\;,
}
where the derivative of the $\chi^2$ has been evaluated at $\vek{p^{\rm b}}$. Provided that the biased Fisher matrix (\ref{biasedfisher}) has an inverse, too, one can solve for the bias and obtain
\eq{
\label{eq:biasgeneral}
b_i = \sum_j \br{{F'}^{-1}}_{\mu\nu} \sum_{\alpha,\beta} P_\alpha^{\rm sys} \br{C_P^{-1}}_{\alpha\beta} \left. \frac{\partial P_\beta}{\partial p_j} \right|_{\rm f}\;.
}
If one assumes that the systematic is small such that the second term in (\ref{biasedfisher}) becomes subdominant, (\ref{eq:biasgeneral}) reproduces the known bias formula (\ref{eq:bias}).

In summary, the differences in employing the exact likelihood/ $\chi^2$ formalism (\ref{eq:biasedchisquare}) or the Fisher matrix approach (\ref{eq:fisherfinal},\,\ref{eq:bias}) can be reduced to cutting the Taylor expansion in (\ref{eq:taylorchisquare}) after the second order in $\vek{p}$, and dropping the second term in (\ref{biasedfisher}). Both approximations are fair if the amplitude of the systematic and the bias it produces are sufficiently small. 

To quantify the validity of these approximations in the context of this work we create a cosmic shear tomography survey with $N_z=10$ redshift bins without photometric redshift errors. The GI signal is calculated via the linear intrinsic alignment model, with a free overall scaling of $A_{\rm sys}$ to control the amplitude of the systematic. The original GI model corresponds to $A_{\rm sys}=1$. We use $\Omega_{\rm m}$ as the only parameter to be constrained, setting a fiducial value of 0.4 for this exemplary analysis. Thereby, as the GI signal biases $\Omega_{\rm m}$ low, we allow for large biases in a range of still reasonable parameter values. To achieve a suitable magnitude of statistical errors, the survey size is set to $1\,{\rm deg}^2$ and $100\,{\rm deg}^2$, respectively, the remaining parameters kept at the values given in Sect.$\,$\ref{sec:modelling}. The exact errors are calculated via (\ref{eq:biasedchisquare}) on a grid in parameter space with steps of $10^{-4}$ between $\Omega_{\rm m}=0.1$ and $\Omega_{\rm m}=0.5$. While the minimum $\chi^2$ is simply read off the grid values, the $1\sigma$-errors are computed by linear interpolation on the grid, with $\Delta \chi^2 \approx 1$ from the minimum for one degree of freedom.
 
We define the ratios
\eq{
\label{eq:ratioscompare}
{r'}_\sigma \equiv \frac{\sigma_{\chi^2}}{\sigma_{\rm F}}\;; ~~~~~ {r'}_b \equiv \frac{b_{\chi^2}}{b_{\rm F}}\;,
}
where $\sigma_{\chi^2}$ denotes the statistical error on $\Omega_{\rm m}$ obtained by the likelihood calculation, and where $\sigma_{\rm F}$ is the statistical error resulting from the computation of the Fisher matrix. Likewise definitions hold for the bias $b_{\chi^2}$ and $b_{\rm F}$. In Fig.$\,$\ref{fig:fishlikecomp} the ratios ${r'}_\sigma$ and ${r'}_b$ are plotted as a function of $A_{\rm sys}$. Apart from uncertainties due to the finite grid resolution the results for both survey sizes agree very well, but since the bias does not depend on the survey size $A$, and $\sigma \propto 1/\sqrt{A}$, the ratios of bias over statistical error differ by a factor of 10. Thus, the limits within which the bias formalism yields accurate results do not depend on this ratio. Instead, the deviations from the exact $\chi^2$ results are a function of the amplitude of the systematic with respect to the original signal. 

For $A_{\rm sys}=1$, i.e. the default GI signal, we find a deviation of the bias obtained by the Fisher matrix formalism of only $2.4\,\%$, despite the strong systematic. The true bias is less than $10\,\%$ larger throughout, even for a very large systematic that dominates the signal by far. In the analysis considered here, both the curvature of the GG power spectrum and the systematic power spectrum are negative, so that the second term in (\ref{biasedfisher}) should in general be negative, too. Consequently, $F'<F$, causing (\ref{eq:biasgeneral}) to produce larger biases than (\ref{eq:bias}), which is evident in Fig.$\,$\ref{fig:fishlikecomp}.

If the amplitude of the systematic increases, the second term in (\ref{biasedfisher}) becomes more important, thereby leading to a scaling of the bias with less than $A_{\rm sys}$ in (\ref{eq:biasgeneral}). Hence, the ratio of biases can curb down for large $A_{\rm sys}$ because the bias, as computed from (\ref{eq:bias}), continues to scale with $A_{\rm sys}$, an effect which is also seen in the figure. A similar behavior may be expected from the inclusion of the third-order in (\ref{eq:taylorchisquare}) as it leads to a term with bias squared in (\ref{eq:chisquareextremum}), thereby placing the term scaling with $P^{\rm sys}$ under a square root when solving for $\vek{b}$.

In the presence of a bias a more accurate way to obtain statistical errors than using the original Fisher matrix would be via $F'$. As opposed to the Fisher matrix formalism, the statistical errors become dependent on the systematic. Inspecting (\ref{biasedfisher}), errors scale linearly with $A_{\rm sys}$ and should increase because of $F'<F$. Again Fig.$\,$\ref{fig:fishlikecomp} demonstrates that this holds true to good approximation, yielding already a $8\,\%$ effect at $A_{\rm sys}=1$. Downscaling the systematic to $A_{\rm sys}=0.2$, the bias formalism should produce results that are very close to the full likelihood calculation, even for the full set of cosmological parameters.

\end{appendix}

\end{document}